\documentclass[prx,twocolumn,superscriptaddress,showpacs,floatfix,longbibliography,10pt]{revtex4-2}

\usepackage{color}
\usepackage{amsfonts}
\usepackage{subfigure}
\usepackage{amsmath}
\usepackage{booktabs}
\usepackage{amssymb}
\usepackage{amsbsy} 
\usepackage{epsfig}
\usepackage{graphicx}
\usepackage{epstopdf}
\usepackage{float}
\usepackage{hyperref}
\usepackage[dvipsnames]{xcolor}
\hypersetup{colorlinks=true, linkcolor=blue, anchorcolor=blue, citecolor=blue,urlcolor=blue}

\def\be{\begin{equation}} \def\ee{\end{equation}}
\def\bea{\begin{eqnarray}} \def\eea{\end{eqnarray}}
\def\nn{\nonumber}

\newcommand{\bra}[1]{\langle#1|}
\newcommand{\ket}[1]{|#1\rangle}

\def\ra{\rangle}
\def\la{\langle}

\begin{document}

\title{Non-Bloch edge dynamics of non-Hermitian lattices}

\author{Wen-Tan Xue}
\affiliation{ Institute for
	Advanced Study, Tsinghua University, Beijing,  100084, China }
\affiliation{ Department of Physics, National University of Singapore, Singapore 117542, Singapore }

\author{Fei Song}
\affiliation{ Institute for
	Advanced Study, Tsinghua University, Beijing,  100084, China }
 \affiliation{Kavli Institute for Theoretical Sciences, Chinese Academy of Sciences, 100190 Beijing, China }

 \author{Yu-Min Hu}
 \affiliation{ Institute for
	Advanced Study, Tsinghua University, Beijing,  100084, China }
 \affiliation{Max Planck Institute for the Physics of Complex Systems, N\"{o}thnitzer Str. 38, 01187 Dresden, Germany}

\author{Zhong Wang}
 \altaffiliation{ wangzhongemail@tsinghua.edu.cn }
\affiliation{ Institute for
	Advanced Study, Tsinghua University, Beijing,  100084, China }

\begin{abstract}

The non-Hermitian skin effect, i.e., the localization of nominally bulk modes, not only drastically reshapes the spectral properties of non-Hermitian systems, but also dramatically modifies the real-time dynamics therein. Here we investigate the time evolution of waves (or quantum-mechanical particles) initialized around the edge of non-Hermitian lattices. The non-Hermitian skin effect tends to localize the wave to the edge, meaning that the real-time dynamics differs from the Bloch-theory picture. We focus on the long-time decay or growth rate of wave function, which is quantified by the Lyapunov exponents. These exponents can be obtained from the saddle points in the complex momentum space. We propose an efficient yet unambiguous criterion for identifying the dominant saddle point that determines the Lyapunov exponents. Our criterion  can be precisely formulated in terms of a mathematical concept known as the Lefschetz thimble. Counterintuitively, the seemingly natural criterion based on the imaginary part of the energy fails. Our work  provides a coherent theory for characterizing the real-time edge dynamics of non-Hermitian lattices. Our predictions are testable in various non-Hermitian physical platforms.

\end{abstract}
\maketitle

\section{Introduction}
With the advancing development in non-Hermitian physics, it has been revealed that boundaries markedly influence the characteristics of non-Hermitian (NH) systems. For instance, under open boundary conditions (OBC), the eigenstates may exhibit the non-Hermitian skin effect (NHSE)\cite{Yao2018,NHchern2018,Kunst2018,Lee2019,Martinez2018,xiao2020,kawabata2020,Helbig2020,Ashida2020,Zhang2020,gohsrich2024,Wang2024,amoeba2024}, leading to a dramatic difference between the OBC and periodic boundary condition (PBC) spectra and profoundly altering topological phenomena such as the bulk-boundary correspondence \cite{Yao2018,NHchern2018,Song2019real,Murakami2019,Okuma2020,Ding2022}. Moreover, these boundary effects also yield novel phenomena in wavefunction dynamics, such as unidirectional amplification \cite{McDonald2018,Wanjura2020,Wentan2021}, quantum sensing enhancement \cite{McDonald2020,Budich2020,Lau2018,Koch2022}, and distinctive entanglement dynamics \cite{kawa2023entanglement,chen2020,Gopa2021,Turkeshi2023,Jian2021} highlighting the diverse consequences that non-Hermiticity introduces into quantum systems \cite{Song2019,liu2020helical,edgeburst2022,haga2021liouvillian,Xiao2024,Longhi2022healing,Hu2023burst,yi2020,XuePeng2024}.

In this study, we investigate the long-time evolution of waves (or quantum-mechanical particles) initialized around the edge of one-dimensional non-Hermitian systems exhibiting NHSE. We focus on two kinds of Lyapunov exponents that quantify the long-time growth or decay of wave function amplitude. The first kind characterizes the temporal evolution of the wavefunction amplitude at the initial site $x_0$, denoted by $\lambda$ and $\mu$ (each relevant to a different timescale, as discussed later). The second kind captures the evolution of the total wavefunction norm across all sites, denoted by $\lambda_\text{tot}$ and $\mu_\text{tot}$. These exponents are closely related to the saddle points of the Bloch spectrum $E(k)$. Previous studies have also shown that saddle points play a crucial role in various aspects of non-Hermitian systems, such as wavefunction dynamics \cite{Longhi2019probing}, frequency response \cite{Zhou2024}, and PT-symmetry breaking transitions \cite{Hu2024}. However, a reliable method for identifying the dominant saddle point that controls the non-Hermitian dynamics has been lacking. To address this major gap, in this work, we propose an efficient criterion for identifying the dominant saddle point among multiple candidates. Our approach is related to a mathematical concept known as the Lefschetz thimble \cite{berry1991,howls1997,lefschetz2012,pham1983}. Specifically, our criterion involves counting the intersection points between the Brillouin zone (BZ)---or equivalently, the generalized Brillouin zone (GBZ) \cite{Yao2018,NHchern2018,Murakami2019})---and the steepest ascent path attached to each saddle point, as detailed further in Section II. B. This criterion is one of the central results of this work.

The reason for using two symbols $\lambda$ and $\mu$ ($\lambda_\text{tot}$ and $\mu_\text{tot}$) to describe the local (global) Lyapunov exponent is that the value of the Lyapunov exponent varies across different time scales: $t\in[0, t_c]$ and $t\in[t_c, \infty)$, with a crossover timescale $t_c$ determined by the system size and group velocity. In the short-time regime $t\in[0,t_c]$, where the evolution is predominantly influenced by the bulk properties, the Lyapunov exponents $\lambda$ and $\lambda_\text{tot}$ are determined by the dominant saddle point. Conversely, in the long-time regime $t\in[t_c,+\infty)$, boundary effects become significant, and the Lyapunov exponents $\mu$ and $\mu_\text{tot}$ correspond to the largest imaginary part of the OBC spectrum. This behavior illustrates how the presence of edges can modify the dynamical properties of non-Hermitian systems.

In addition, we discover two interesting phenomena : (1) with the existence of boundary, the wave peak may exhibit an opposite moving direction with the NHSE, which is dubbed ``non-sticky effect". (2) a large number of OBC eigenstates (rather than the semi-infinite boundary conditions (SIBC) eigenstates introduced in \cite{Longhi2022healing}) in non-Hermitian systems can recover themselves after being significantly perturbed by a large potential—a phenomenon known as the self-healing effect. Both phenomena can be explained by our Lefschetz-based analysis of $\lambda_\text{tot}$.

\section{Dynamics on the initial edge}
\begin{figure}[H]
\includegraphics[width=9cm, height=3cm]{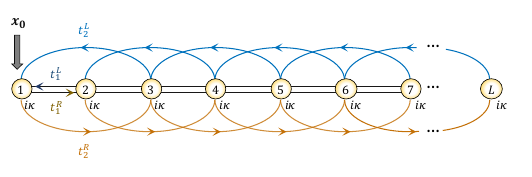}
\caption{Schematic representation of the non-Hermitian Hamiltonian in Eq~\eqref{hk}, with $x_0$ denoting the particle's initial position.}
\label{model}
\end{figure}

Although our method is generally applicable to a generic one-dimensional non-Hermitian systems,  we here focus on a simple single-band model with NHSE for concreteness. This model, shown in Fig.~\ref{model}, describes a lossy lattice with asymmetric nearest- and next-nearest-neighbor hoppings. Its Bloch Hamiltonian is given by:
\be
h(k)=t_1^Le^{ik}+t_1^Re^{-ik}+t_2^Le^{2ik}+t_2^Re^{-2ik}-i\kappa.
\label{hk}
\ee
For this model, the energy spectrum $E(k)$ is equal to $h(k)$, so we use $h(k)$ in place of $E(k)$ throughout the article. Moreover, for simplicity, we focus on parameter regimes where the skin modes are localized exclusively at the left edge. In particular, we refer to the case where all skin modes are trapped at only one side of the system as unidirectional NHSE.
 
\subsection{Lyapunov exponents across  two time scales}

\begin{figure}
\includegraphics[width=8.4cm, height=12cm]{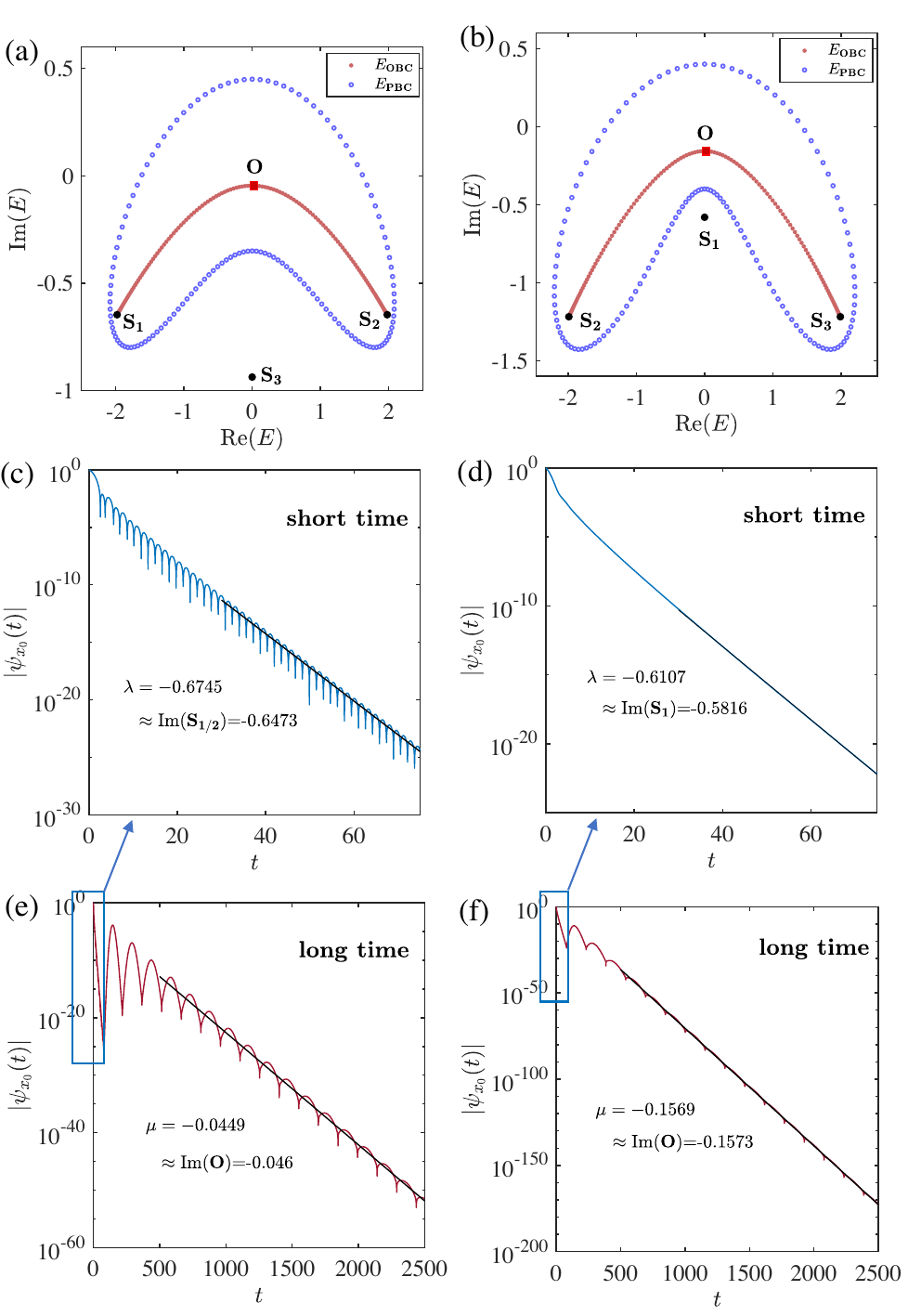}
\caption{Energy spectra and numerical fitting of Lyapunov exponents $\lambda$ and $\mu$. (a,b) OBC (red dots) and PBC (blue circles) energy spectra with saddle points $S_{1,2,3}$ marked explicitly ($S_4$ lies outside the plotted range). Parameters: (a,c,e) $t_1^L=1.2i, t_1^R=-0.8i, t_2^L=0.35i, t_2^R=0.05i, \kappa=0.35$; (b,d,f) $t_1^L=1.2i, t_1^R=-0.8i, t_2^L=0.6i, t_2^R=0.1i,\kappa=0.7$, both with system size $L=140$. Red squares (O) indicate the eigenenergies with the largest imaginary part of the OBC spectrum. (c,d) Short-time numerical fitting of $|\psi_{x_0}(t)|$. (e,f) Long-time numerical fitting of $|\psi_{x_0}(t)|$.}
\label{fig2}
\end{figure}

We now consider the motion of a wave-packet on a one-dimensional open chain governed by the Bloch Hamiltonian in Eq~\eqref{hk}. The initial state is a delta function localized at the leftmost site, i.e., $\psi_x(t=0)=\la x\ket{x_0=1}=\delta_{x,1}$. In this non-Hermitian system, the amplitude at the initial site, $|\psi_{x_0}(t)|=|\bra{x_0}e^{-iHt}\ket{x_0}|$, can grow or decay exponentially over time. Fig. \ref{fig2} illustrates the typical time evolution of $|\psi_{x_0}(t)|$ for two different parameter sets. Interestingly, in both cases, $|\psi_{x_0}(t)|$ exhibits different Lyapunov exponents over two time scales, as shown below:
\bea
\left|\psi_{x_0}(t)\right|\sim \begin{cases} e^{\lambda t}
, & t<t_c\\
e^{\mu t}, & t\gg t_c,
\end{cases}
\eea
where $\lambda$ and $\mu$ are the local Lyapunov exponents. Moreover, based on our numerical observations, the crossover timescale $t_c$ that divides the short-time and long-time regimes is proportional to the system length. It will be thoroughly analyzed at the end of this paper. Now, our main focus is to derive the general characteristics of the two local Lyapunov exponents. In Fig.~\ref{fig2}, by performing a numerical fit of $|\psi_{x_0}(t)|$ on a logarithmic scale, we get:
\bea
\ln\left|\psi_{x_0}(t)\right|=\begin{cases}
\lambda t+c_1, &t< t_c\\
\mu t+c_2, & t\gg t_c.
\end{cases}
\eea
We find $\lambda \approx -0.6745(-0.6107)$ for the short-time regime ($t<t_c$), as shown in Figs.~\ref{fig2}(c) and \ref{fig2}(d). In the long- time regime ($t\gg t_c$), as shown in Figs.~\ref{fig2}(e) and ~\ref{fig2}(f), we observe that the exponent increases to $\mu\approx-0.0449(-0.1569)$. This value matches the largest imaginary part of the
OBC spectrum, indicated by the point $O$ (red squares) in Figs.~\ref{fig2}(a,b). This can be directly demonstrated as follows:
\bea \lim_{t\rightarrow \infty}  \left|\psi_{x_0}(t)\right| &=& \left|\langle x_0| e^{-iHt}|x_0\rangle\right|\nn\\
 &=& \left|\sum_n e^{-iE_n t}\langle x_0|\psi_{n}^R\rangle\langle \psi_{n}^L|x_0\rangle\right|\nn\\
 &\approx & \max_n \left| e^{-iE_n t}\langle x_0|\psi_{n}^R\rangle\langle \psi_{n}^L|x_0\rangle \right|\nn\\
&\sim& e^{\text{Im}(O)t}, 
 \label{psix0} \eea
where $\ket{\psi_n^{R}}$ and $\ket{\psi_n^{L}}$ are the right and left eigenstates of $H$, respectively. In the third step, we retain only the dominant contribution---specifically, the term corresponding to the eigenstate at point $O$---since for large $t$ the contributions from other eigenstates satisfy $|e^{-iE_nt}|\ll e^{\text{Im}(O)t}$. Thus, in the long-time limit the Lyapunov exponent is given by 
\be
\mu=\text{Im}(O)=\max\{\text{Im} E_n |E_n\in\text{OBC spectrum}\},
\ee
which is verified in Figs.~\ref{fig2}(e) and \ref{fig2}(f).

Conversely, during $t\in [0 ,t_c]$, the eigenstates with energies close to the point $O$ are not yet dominant in the summation in Eq~\eqref{psix0}. Instead, the contributions from all $n$ eigenstates interfere, effectively reducing the magnitude of the overall sum, leading to a smaller Lyapunov exponent $\lambda$, as observed in Fig.~\ref{fig2}. Importantly, $\lambda$ is found to be associated with the saddle-point energies of the Bloch Hamiltonian $h(k)$, denoted by $S_\sigma =h(k^s_\sigma)$ ($\sigma=1,2,3,4$), as illustrated in Figs.~\ref{fig2} (a) and \ref{fig2}(b). Here, the saddle points $k_\sigma^s$ are defined by
\be 
\left. \frac{dh(k)}{dk}\right|_{k=k^s_\sigma}=0.
\ee
Notably, although the two endpoints of the OBC spectrum are saddle points, not all saddle points lie on it; for instance, the dominant saddle point
$S_1$ in Fig.~\ref{fig2}(b), lies outside the OBC spectrum. In fact, a recent study \cite{Hu2024} has shown that saddle points generally reside on the auxiliary generalized Brillouin zone (aGBZ) \cite{aGBZ2020} rather than the GBZ itself.

To understand how the exponent $\lambda$ is determined by the saddle points, we first express the OBC Hamiltonian $H$ in terms of the GBZ: 
 \bea
\left|\psi_{x_0}(t)\right|_{t<t_c}
&=&\left|\sum_n e^{-iE_nt}\la x_0|\psi_n^R\ra\la\psi_n^L|x_0\ra\right|\nn\\
&\approx&\left|\sum_{k_n\in \text{GBZ}}e^{-ih(k_n)t}\la x_0|\psi_{k_n}^R\ra\la\psi_{k_n}^L|x_0\ra\right|\nn\\
&\approx&\left|\frac{1}{L}\sum_{k_n\in \text{GBZ}}e^{-ih(k_n)t}e^{ik_nx_0}R(k_n)e^{-ik_nx_0}L(k_n)\right|\nn\\
&\approx&\left|\frac{1}{L}\sum_{k_n\in \text{GBZ}}e^{-ih(k_n)t}g(k_n)\right|,
\label{eq5}
\eea
where we have used $\la x_0|\psi_{k_n}^R\ra=e^{ik_nx_0}R(k_n)/\sqrt{L}$, $\la \psi_{k_n}^L|x_0\ra=e^{-ik_nx_0}L(k_n)/\sqrt{L}$, 
and defined $g(k_n)=R(k_n)L(k_n)$.
During the short time interval $t<t_c$, the wave packet remains far from the right end, allowing us to equivalently treat the system as an infinite long lattice and approximate the discrete momentum $k_n$ as continuous. Thus, the summation over $k_n$ can be replaced by an integral over the GBZ. Furthermore, as demonstrated in the Appendix~\ref{app:B}, the integrals over the BZ and GBZ yield the same value. This leads to:
 \bea
\left|\psi_{x_0}(t)\right|_{t<t_c}
&\approx&\frac{1}{2\pi}\left|\int_{\text{GBZ}}e^{-ih(k)t}g(k) dk\right|\nn\\
&\approx&\frac{1}{2\pi}\left|\int_{\text{BZ}}e^{-ih(k)t}g(k) dk\right|,
\label{eqGBZBZ}
\eea
where we have substituted $dk\sim 2\pi/L$ in the first step.
To evaluate this complex integral, a commonly followed approach is the method of steepest descents \cite{steepest}. The core idea is that the original integral path can be deformed into a new path $\mathcal{C}'$ in the complex $k$-plane, satisfying two conditions: (1) the new path $\mathcal{C}'$ passes through the saddle point of $h(k)$; and (2) the real part of $h(k)$, Re$[h(k)]$, remains constant along $\mathcal{C}'$. Under this two conditions, the integral transforms as:
\bea
\int_{\text{BZ}} e^{-ih(k)t}g(k)dk &=&\int_{\mathcal{C}'} e^{-ih(k)t}g(k)dk\nn\\
&=& e^{-i\text{Re}[h]t} \int_{\mathcal{C}'} e^{\text{Im}[h(k)]t}g(k)dk,
\label{eq6}
\eea
where along the path $\mathcal{C}'$, $\text{Re}[h]$ is a $k$-independent constant. 

As demonstrated in the Appendix~\ref{app:A}, since Re$[h(k)]$ remains constant along the path $\mathcal{C}'$, this path must correspond to the steepest ascent or descent direction of $\text{Im}[h(k)]$ from the saddle point. To ensure saddle point having the most dominant contribution, we select the steepest descent path as the new integration contour, $\mathcal{C}'$. Along $\mathcal{C}'$, $\text{Im}[h(k)]$ is a Morse function \cite{arnold2012,milnor1963,matsumoto2002}, reaching its the maximum value at the saddle point $k^s$. Using the complex Morse Lemma\cite{milnor1963,matsumoto2002}, we can simplify the integral by keeping the contributions from the neighborhood of $k^s$, resulting in: 
\bea
\int_{\mathcal{C}'} e^{\text{Im}[h(k)]t}g(k)dk&\approx & e^{\text{Im}(S)t}g(k^s)+O(\Delta k^2)\nn\\
&\sim&  e^{\text{Im}(S)t}
\label{eq10}
\eea
where the saddle point $S=h(k^s)$ and $\Delta k=k-k^s$ is tiny. This result indicates that the Lyapunov exponent $\lambda$ is determined by the imaginary part of the saddle point.

However, a key issue arises: for the Hamiltonian in Eq~\eqref{hk} and more general non-Hermitian systems, $h(k)$ can exhibit multiple saddle points. Which of these saddle points determines the Lyapunov exponent $\lambda$? Based on the numerical results in Fig.~\ref{fig2}, one may be tempted to think that the saddle point whose energy has the largest imaginary part dominates $\lambda$. However, this conjecture is not always valid. By employing the Lefschetz thimble method, we establish an unambiguous general criterion for identifying the truly dominant saddle point, as explained below.  

\begin{figure*}
\centering
\includegraphics[width=17.4cm, height=4cm]{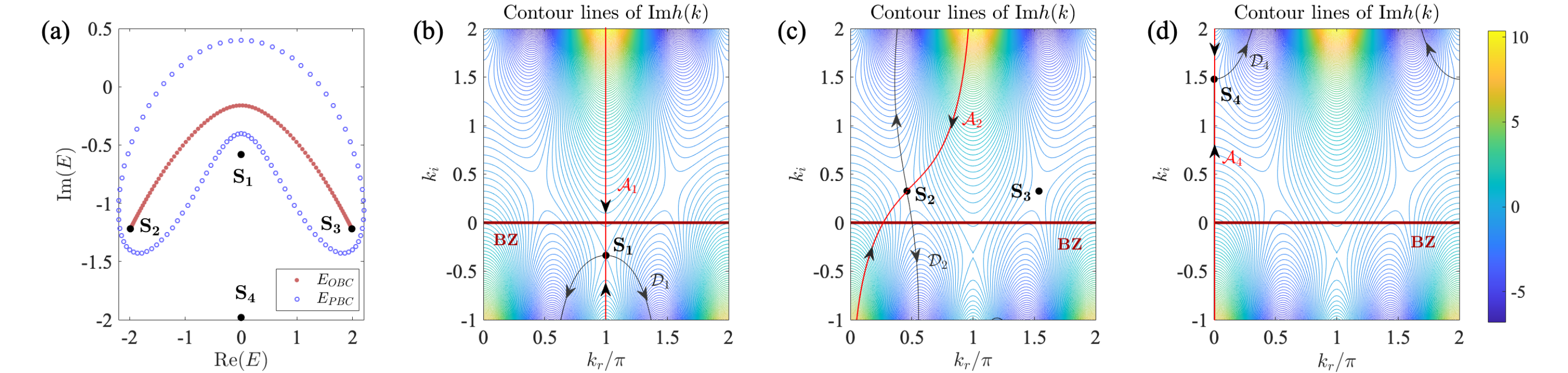}
\includegraphics[width=17.4cm, height=4cm]{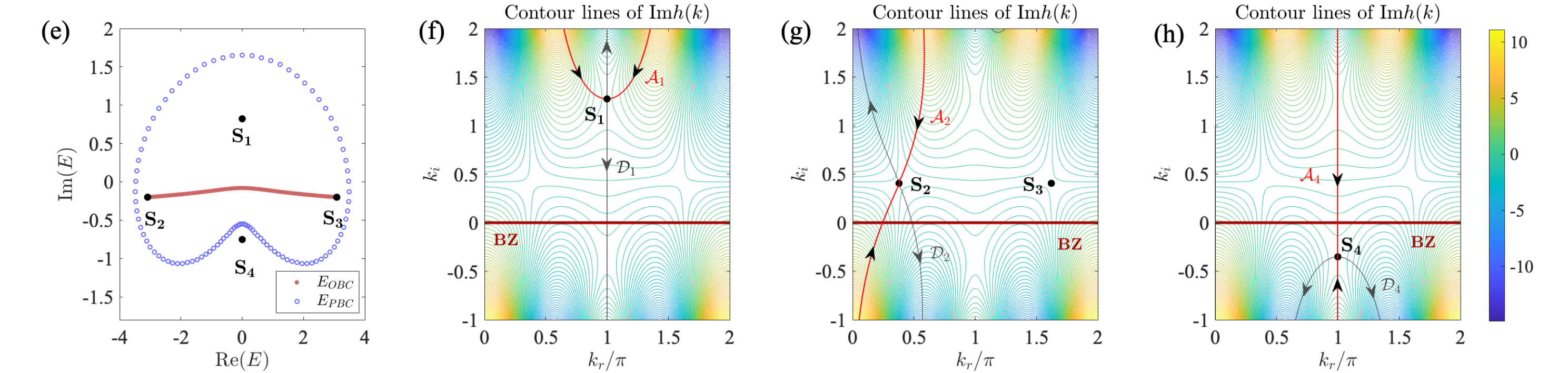}
\caption{Analysis of the dominant saddle points in two representative examples. Panels (a-d) correspond to the parameter set in Fig.~\ref{fig2}(b), while panels (e-h) correspond to the parameters from Fig.~\ref{fig4}(e). The red curves labeled BZ in (b-d) and (f-h) represent the Brillouin zone, and the black arrows illustrate the downward flow of $\text{Im}[h(k)]$. Curves $\mathcal{A}_\sigma$ and $\mathcal{D}_\sigma$ denote the steepest ascent and descent paths associated with each saddle point, respectively. Note that each ascent curve,
except $\mathcal{A}_1$ in panel (f), intersects the BZ exactly once.}
\label{fig3}
\end{figure*}

\subsection{Identifying the dominant saddle point through Lefschetz thimble method}
\begin{figure*}
\includegraphics[width=16.8cm, height=8cm]{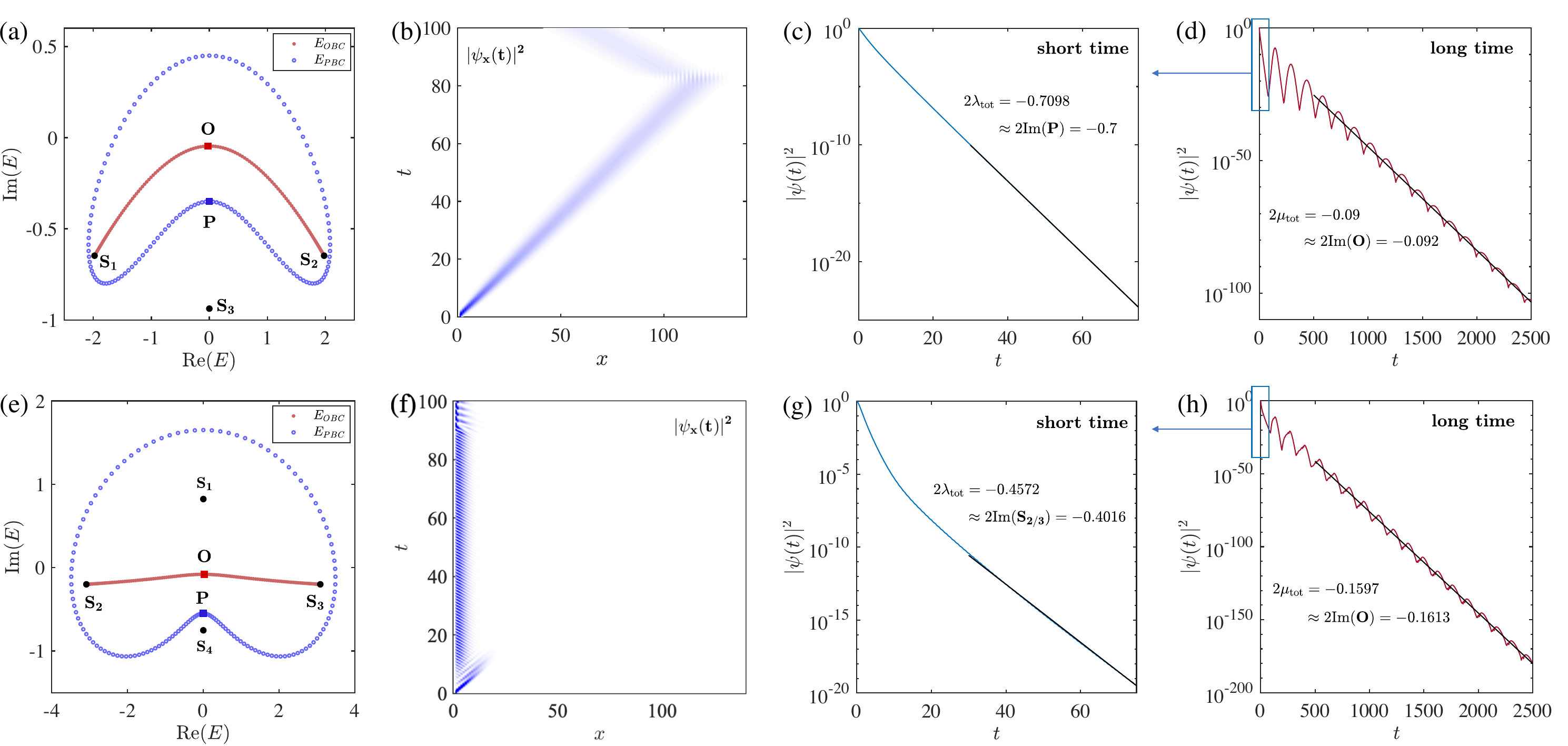}
\caption{Two distinct evolution behaviors of the wave peak with system size $L=140$; (a-d) Results with the same parameter set as in Fig.~\ref{fig2}(a). (e-h) Results with parameters: $t_1^L=2.05i, t_1^R=-0.95i, t_2^L=0.85i, t_2^R=-0.15i, \kappa=0.15$. (a, e) Energy spectra under both OBC and PBC, where points $P$ mark the local maximum of the imaginary part of the PBC spectrum. (b, f) Time evolution of the wavefunction $|\psi_x(t)|^2$ in the short time regime, with renormalization applied at each time step. (c, g) Numerical fitting of the Lyapunov exponent $\lambda_\text{tot}$ for short-time regime. (d)(h) Numerical fitting of $\mu_\text{tot}$ for the long-time regime.}
\label{fig4}
\end{figure*}

In this section, we provide two representative examples (Fig.~\ref{fig3}) to demonstrate how the Lefschetz thimble method can be used to identify the dominant saddle point, denoted by $S_d$ , in the short-time dynamics. In the first example (Figs.~\ref{fig3}(a-d)), the dominant saddle point $S_d$ is given by $S_1$, which has the largest imaginary energy among all saddle points, as verified in Fig.~\ref{fig2}(d). In contrast, in the second example (Fig.~\ref{fig3}(e-h)), $S_d$ corresponds $S_{2}$ (or $S_3$), rather than $S_1$, as confirmed in Fig.~\ref{fig4}(g). This implies that saddle point $S_1$ makes no contribution to the integral in Eq~\eqref{eq10}, thereby clearly disproving the earlier conjecture that the dominant saddle point always possesses the largest imaginary energy.

To accurately determine $S_d$ and thus the Lyapunov exponent $\lambda$, we must identify whether each saddle point contributes to the integral in Eq~\eqref{eq10}. As elaborated later, the conclusion is that the contribution of a given saddle point $S_\sigma$ is characterized by the coefficient $n_\sigma$, defined as follows:
\bea
n_\sigma &=&\langle \text{BZ},\mathcal{A}_\sigma\rangle=\begin{cases}
0, &\text{even intersection points}\\
\pm 1, &\text{odd intersection points}
\end{cases},
\label{n_sigma}
\eea
where $\mathcal{A}_\sigma$ is the steepest ascent path associated with $S_\sigma$, and $\la \text{BZ},\mathcal{A}_\sigma\ra$ denotes the intersection pairing between the BZ and the curve $\mathcal{A}_\sigma$. A saddle point $S_\sigma$ contributes if $n_\sigma=\pm 1$ and does not contribute if $n_\sigma=0$. For example, in  Fig.~\ref{fig3}(b), the BZ intersects $\mathcal{A}_1$ exactly once; thus $n_1=1$, and $S_1$ is contributive. Similarly, $S_2,S_3$ and $S_4$ are also contributive, as shown in Figs.~\ref{fig3}(c) and (d). Among these contributive saddle points, Im($S_1$) is the largest, resulting in the dominant short-time behavior $|\psi_{x_0}(t)|_{t<t_c}\sim e^{\text{Im}(S_1)t}$. In contrast, for $S_1$ in Figs.~\ref{fig3}(e,f), we have $n_1=\la\text{BZ},\mathcal{A}_1\ra=0$, meaning it does not contribute. Consequently, $S_d$ becomes $S_2$ (or equivalently $S_3$, since Im($S_2$)=Im($S_3$)), both of which satisfy $n_{2,3}=1$.

Let's now return to the origin of the expression for $n_\sigma$ in Eq~\eqref{n_sigma}. According to the steepest descent method and related studies on path integrals in Chern--Simons theory \cite{witten2011,Eastham1985,pham1983,berry1991,howls1997}, when multiple saddle points are present, the deformed integration contour $\mathcal{C'}$ introduced in Section I.A can be expressed as a linear combination of Lefschetz thimbles $\mathcal{D}_\sigma$:
\be
\mathcal{C'}=\sum_\sigma n_\sigma \mathcal{D}_\sigma.
\label{eqS2}
\ee
Each Lefschetz thimble is the steepest descent path of Im$[h(k)]$ attached to a saddle point $k^s_\sigma$. Substituting $\mathcal{C}'$ into Eq~\eqref{eqGBZBZ} and \eqref{eq6}, we obtain
\bea
\left|\psi_{x_0}(t)\right|_{t<t_c}&\approx&\left|\sum_\sigma \frac{n_\sigma}{2\pi} \int_{\mathcal{D}_\sigma}g(k)e^{-i\text{Re}[h(k)]t}e^{\text{Im}[h(k)]t}dk\right|\nn\\
&\approx& \left|\sum_\sigma \frac{n_\sigma}{2\pi} e^{-i\text{Re}(S_\sigma)t}g(k^s_\sigma)e^{\text{Im}(S_\sigma)t}\right|.
\label{eq11}
\eea
According to Morse theory, $n_\sigma$ is determined by the intersection pairing between the deformed contour $\mathcal{C'}$ (equivalent to the original contour BZ) and the steepest ascent path $\mathcal{A}_\sigma$ of $\text{Im}[h(k)]$. Considering the orientation at each intersection, we obtain the Eq~\eqref{n_sigma}:
\be
n_\sigma=\langle \text{BZ},\mathcal{A}_\sigma\rangle=\sum_i\text{sgn}(p_i),
\ee
where $p_i$ represents the $i$-th intersection point, and $\text{sgn}(p_i)$ indicates its orientation. In particular, if there is an even number of intersection points, $\mathcal{A}_\sigma$ must intersect BZ from opposite directions, resulting in a zero net contribution ($n_\sigma=0$). 

Therefore, Eq~\eqref{eq11} can be approximated as 
\be
\left|\psi_{x_0}(t)\right|_{t<t_c}\sim e^{\text{Im}(S_d)t}\sim e^{\lambda t},
\ee
with 
\be
\lambda=\text{Im}(S_d)=\text{max}_{n_\sigma\ne 0}\text{Im}(S_\sigma),
\ee
i.e., the dominant saddle point $S_d$ is selected as the one with the largest imaginary part among all saddle points for which $n_\sigma\ne 0$.

We finally conclude that for the evolution of $\psi_{x_0}(t)$, the local Lyapunov exponent is given by 
\bea 
 \begin{cases} \lambda=\operatorname{Im}(S_d)=\text{max}_{n_\sigma\ne 0}\text{Im}(S_\sigma), & t < t_c, \\
\mu=\operatorname{Im}(O)=\text{max}\left[\text{Im}(E_\text{OBC})\right], & t \gg t_c, \end{cases} 
\label{lambda&mu}
\eea
where $S_d$ is the dominant saddle point of $h(k)$ and $O$  denotes the eigenvalue with the largest imaginary part in the OBC spectrum. Eq. (\ref{lambda&mu}) is a central result of this work. The method used to determine $t_c$ will be described in a later section.

\section{Dynamics of the wavefunction norm}

Having established the general behavior of $\psi_{x_0}(t)$, we now turn
to the evolution of the wavefunction norm across all sites and to the associated Lyapunov exponents, $\lambda_\text{tot}$ and $\mu_\text{tot}$, defined by
 \be
\left|\psi(t)\right|=\sqrt{\sum_x|\psi(x,t)|^2}\sim 
\begin{cases}
e^{\lambda_\text{tot}t}, &t<t_c\nn\\
e^{\mu_\text{tot}t}, & t\gg t_c.
\end{cases}
 \ee
In the long time regime $t\in[t_c,+\infty)$, the situation is analogous to the previous analysis of 
$\psi_{x_0}(t)$. Specifically,
\bea
\lim_{t\rightarrow\infty}\left|\psi(t)\right|&=&\sqrt{\sum_x|\langle x|e^{-iHt}|x_0\rangle|^2}\nn\\
&=&\sqrt{\sum_{x,n}e^{2\text{Im}(E_n)t}|\la x|\psi^R_{n}\ra\la\psi^L_{n}|x_0\ra|^2}\nn\\
&\sim & e^{\text{Im}(O)t}.
\label{eq18}
\eea
Hence,
\be
\mu_\text{tot}=\text{Im}(O)=\max\{\text{Im} E_n |E_n\in\text{OBC spectrum}\},
\label{mu_tot}
\ee
which is consistent with the fittings in Figs.~\ref{fig4} (d) and \ref{fig4}(h).

\begin{figure}
\includegraphics[width=8.4cm, height=4cm]{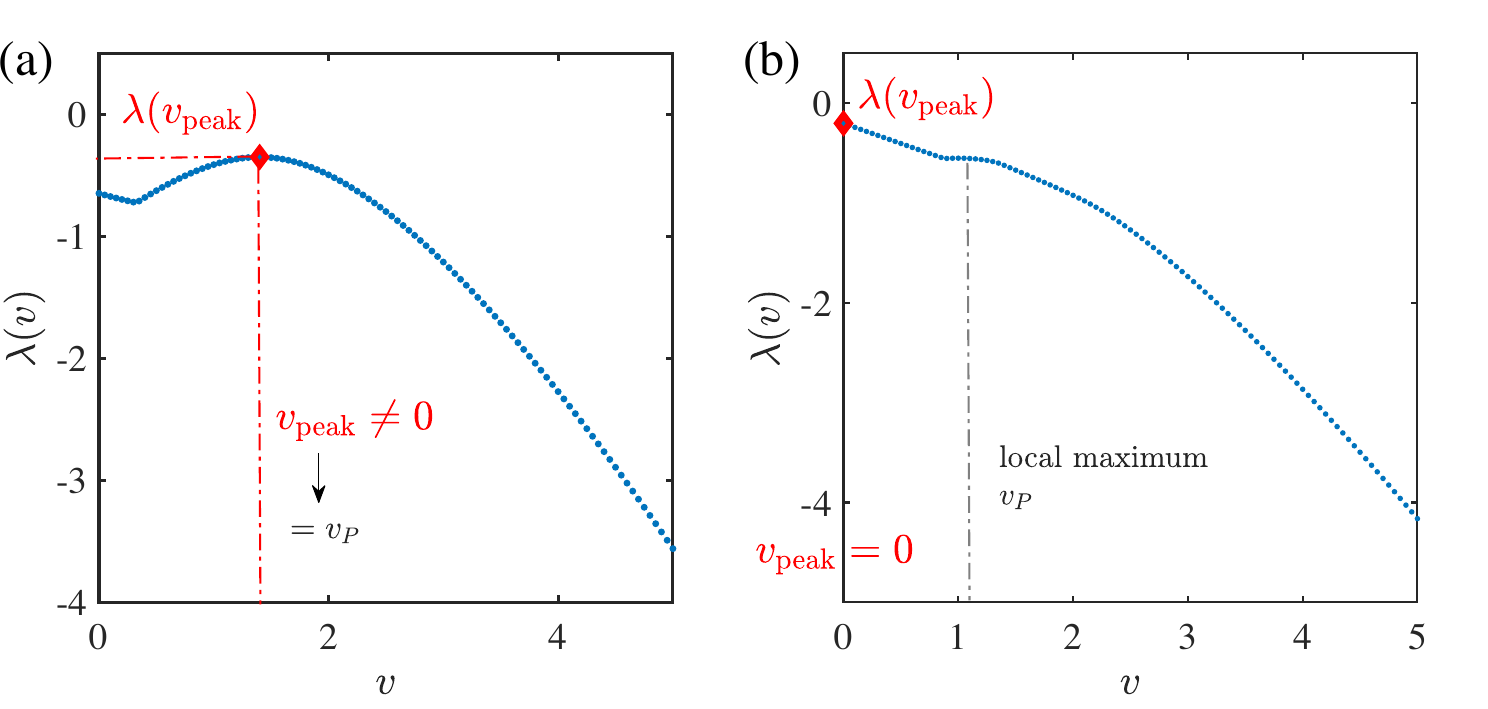}
\caption{Two typical examples of the function $\lambda(v)$, calculated using Eq~\eqref{lambda_v}, where $k^s_d$ corresponds to the dominant saddle point.
(a) Using the same parameters as in Figs.~\ref{fig4}(a-d). (b) Using the same parameters as in Figs.~\ref{fig4}(e-h), where $v_P$ denotes the position of a local maximum. It is derived in the Appendix~\ref{app:C} that $v_P$ corresponds the velocity at point $P$ in Figs.~\ref{fig4}(a) and (e).}
\label{fig5}
\end{figure}

In the short time regime $t<t_c$, we first analyze the amplitude of summand $\la x|e^{-iHt}|x_0\ra$ in a similar manner as in Eqs.~\eqref{eq5} and \eqref{eqGBZBZ}:
\bea
\la x|e^{-iHt}|x_0\ra
&=&\sum_n e^{-iE_nt}\la x_0|\psi_n^R\ra\la\psi_n^L|x_0\ra\nn\\
&\approx&\frac{1}{2\pi}\int_\text{BZ} e^{ik(x-x_0)}e^{-ih(k)t}g(k)dk.
\eea
By defining a drift velocity as $v=\frac{x-x_0}{t}$, , the above equation becomes:
\be
\la x|e^{-iHt}|x_0\ra\approx\frac{1}{2\pi}\int_\text{BZ} e^{-i[h(k)-kv]t}g(k)dk.
\ee
Importantly, the velocity has to fulfill $v\ge 0$ since the wavepacket starts evolving from the leftmost site of the system and all other sites satisfy $x>x_0=1$. Thus, the wavefunction norm in Eq~\eqref{eq18} can be written as
\be
\sum_x|\psi(x,t)|^2\approx\sum_{v}\frac{1}{4\pi^2}\left|\int_\text{BZ} e^{-i[h(k)-kv]t}g(k)dk\right|^2.
\label{sumv}
\ee
Similar to our previous analysis of $|\psi_{x_0}(t)|$ for $t<t_c$, the integral on the RHS of Eq~\eqref{sumv} is determined by the saddle points of the function $h'(k)=h(k)-kv$. These $v$-dependent saddle points are given by 
\be
S_\sigma(v)=h(k^s_\sigma)-k^s_\sigma v,
\ee
where $k^s_\sigma$ satisfies $\frac{dh(k)}{dk}|_{k=k^s_\sigma}=v$, and thus depends on $v$. For a fixed $v$, by applying the Lefschetz thimble method introduced in the previous section, we can identify the dominant saddle point $S_d(v)$. Consequently, Eq~\eqref{sumv} becomes
\bea
\sum_x|\psi(x,t)|^2&\approx&\sum_{v\geq 0} \frac{1}{4\pi^2}e^{2\text{Im}[S_d(v)]t}\left|g(k^s_d)\right|^2\nn\\
&\sim& \sum_{v\geq 0} e^{2\lambda(v)t},
\label{sum_lambda_v}
\eea
where
\be
\lambda(v)=\text{Im}[S_d(v)]=\text{Im}\left[h(k^s_d)-k^s_dv\right]
\label{lambda_v}
\ee
represents the decay rate of the wavefunction amplitude at position $x=vt$. Two typical examples of $\lambda(v)$ are shown in Fig.~\ref{fig5}, where the maximum value occurs at $v=v_\text{peak}$. Physically, $v_\text{peak}$ corresponds to the group velocity of the wave packet peak. 

Moreover, in the summation of Eq~\eqref{sum_lambda_v}, contributions from velocities other than $v_\text{peak}$ are negligible, so that
\be
|\psi(t)|=\sqrt{\sum_x|\psi(x,t)|^2}
\sim e^{\lambda(v_\text{peak})t}.
\label{eq22}
\ee
Thus, the global Lyapunov exponent in the short time regime ($t<t_c$) is given by:
\bea
\lambda_\text{tot}&=&\lambda(v_\text{peak})=\text{max}_{v\ge 0}\lambda(v)\nn\\
&=&\text{max}_{v\ge 0;n_\sigma\ne 0}\text{Im}\left[h(k_\sigma^s)-vk_\sigma^s\right].
\label{lambda_tot}
\eea

\subsection{Two distinct evolutionary behaviors of the wave peak for $t<t_c$}

As shown in Fig.~\ref{fig5}, there are two distinct cases for $\lambda(v_\text{peak})$: $v_\text{peak}=0$ and $v_\text{peak}>0$, which correspond to the different wave-packet evolutions in Figs.~\ref{fig4} (b) and (f). Interestingly, the $v_\text{peak}>0$ scenario in Fig.~\ref{fig4}(b) is counterintuitive, because the rightward propagation of the wavepacket is contrary to the expected leftward propagation induced by the NHSE. This phenomenon is allowed because the boundary constrains the drift velocity to be positive. Conversely, as discovered in Ref. \cite{Longhi2019probing}, the drift direction of the wavepacket placed in a PBC system always aligns with the localization direction of skin modes under OBC, unless the system features bi-polar NHSE. This also means that if the function $\lambda(v)$ depicted in Fig.~\ref{fig5}(a) is extended to the negative-velocity side, a new peak with larger Lyapunov exponent can be found at a certain negative velocity. Therefore, this ``skin-opposite'' propagation is an unexpected dynamical phenomenon unique to OBC systems. In addition to placing the initial state exactly at the edge, this exotic phenomenon can also be observed when the initial position of the wave packet is at a finite distance from the edge (see Fig.~\ref{figS2} in the Appendix~\ref{app:D}).

\begin{table*}[htb]
\caption{Lyapunov exponents for $|\psi_{x_0}(t)|$ and $|\psi(t)|$ under two time scales. The saddle point $k_\sigma^s$  is determined by $\partial_k h(k)|_{k=k_\sigma^s}=0$ for $|\psi_{x_0}(t)|$ in the first line and  $\partial_k h(k)|_{k=k_\sigma^s}=v$ for $|\psi(t)|$ in the second line. For each saddle point $k_\sigma^s$, $n_\sigma$ is its intersection pairing number with BZ or GBZ, which is a key quantity exploited in Lefchetz thimble method. The explicit definition of $n_\sigma$ can be found in Eq.~(\ref{n_sigma}). }
\label{table1}
\centering
\begin{tabular}{c c c}
\toprule
 & \(t < t_c\) & \(t \gg t_c\) \\
\midrule
$|\psi_{x_0}(t)|$ & \begin{tabular}[c]{@{}l@{}}$\ \ \ \ \ \ \lambda=\max_{ n_\sigma\neq 0}\text{Im}[h(k_\sigma^s)]$\\ (e.g. $\lambda=\text{Im}(S_1)$ in Figs.~\ref{fig2}(b,d))\end{tabular} & \begin{tabular}[c]{@{}l@{}}$\ \ \ \ \ \ \ \ \mu=\text{max}[\text{Im}(E_\text{OBC})]$\\ (e.g. $\mu=\text{Im}(O)$ in Figs.~\ref{fig2}(b,f))\end{tabular} \\
\hline
$|\psi(t)|$  & \begin{tabular}[c]{@{}l@{}}$\ \ \ \lambda_\text{tot}=\max_{v\geq 0;n_\sigma\neq 0}\text{Im}[h(k_\sigma^s)-vk_\sigma^s]$\\ \ \ \ \ (e.g. $\lambda_\text{tot}=\text{Im}(P)$ in Figs.~\ref{fig4}(a,c))\end{tabular} & \begin{tabular}[c]{@{}l@{}}$\ \ \ \ \ \ \ \mu_\text{tot}=\text{max}[\text{Im}(E_\text{OBC})]$\\ (e.g. $\mu_\text{tot}=\text{Im}(O)$ in Figs.~\ref{fig4}(a,d))\end{tabular} \\
\bottomrule
\end{tabular}
\end{table*}

For the $v_\text{peak}>0$ case in Figs.~\ref{fig5} (a), $v_\text{peak}$ corresponds to the local maximum of the function $\lambda(v)$. As shown in Fig.~\ref{fig4} (b), the wavepacket propagates from the initial position $x_0=1$ to the opposite edge. In contrast, for the $v_\text{peak}=0$ case in Fig.~\ref{fig5} (b), the local maximum $\lambda(v_P)$ (with $v_P$ given by Eq~\eqref{v_P}) is smaller than $\lambda(v=0)$, causing the wavepacket to be dominated by the component with zero velocity. This results in the wave peak ``sticking" to the left edge, as shown in Fig.~\ref{fig4} (f).
Therefore, by finding the velocity associated with the maximum $\lambda(v_P)$,  we can predict whether the wave peak will move or remain stationary. Physically, this behavior can be interpreted as a competition between two effects: the NHSE that tends to trap the wave packet at the left edge and the moving tendency of the wave packet associated with the velocity dependence of the Lyapunov exponent $\lambda(v)$.

First, we analyze the value of $\lambda(v_\text{peak})$ for the ``sticky" case in Fig.~\ref{fig5}(b). By substituting $v_\text{peak}=0$ into Eqs.~\eqref{lambda_v} and \eqref{eq22}, we obtain
\bea
\left|\psi(t)\right|_{t<t_c}&\sim & e^{\lambda(0)t}\sim e^{\text{Im}[S_d t]},
\eea
where $S_d$ is the dominant saddle point of $h(k)$, same as the one in Eq~\eqref{lambda&mu}. Hence, in this case, the short time Lyapunov exponent is
\be
\lambda_\text{tot}=\lambda(v_\text{peak})=\text{Im}(S_d),
\ee
which is confirmed by the numerical fitting shown in Fig.~\ref{fig4}(g).

For the $v_\text{peak}> 0$ case in Fig.~\ref{fig5}(a), $\lambda(v_\text{peak})$ is a local maximum of the function $\lambda(v)$, satisfying $\frac{d\lambda(v)}{dv}|_{v=v_{\text{peak}}}=0$. From the numerical fitting in Fig.~\ref{fig4}(c), we observe that $\lambda(v_\text{peak})$ no longer equals to $\text{Im}(S_d)$. Instead, it corresponds to the imaginary part of point $P$, a local maximum determined by the imaginary part of the PBC spectrum [Fig.~\ref{fig4} (a)]. Hence, $\lambda(v_\text{peak})$ is given by:
\be
\lambda(v_\text{peak})=\text{Im}(P).
\label{eqpeakImp}
\ee
Here, the point $P=h(k_P)$ belongs to the Bloch spectrum and $k_P\in[0,2\pi)$ is a real momentum. A detailed derivation of Eq~\eqref{eqpeakImp} is provided in the Appendix~\ref{app:C}. It is also shown there that the velocity $v_\text{peak}$ at which $\lambda(v)$ attains its local maximum, is determined by the group velocity at point $P$, i.e.,
\be
v_\text{peak}=v_P=\left. \frac{dh(k)}{dk}\right|_{k=k_P}.
\label{v_P}
\ee
Therefore, for the case with $v_\text{peak}>0$, the  wavefunction norm satisfies:
\bea
|\psi(t)|\sim e^{\lambda_\text{tot}t}=e^{\text{Im}(P)t}.
\eea
From the analysis of these two cases, we conclude that, for our model, the general form of the global Lyapunov exponent (Eq~\eqref{lambda_tot}) becomes 
\be
\lambda_\text{tot}=\text{max}\left[\text{Im}(S_d),\text{Im}(P)\right].
\ee

For a general non-Hermitian model with Bloch Hamiltonian $h(k)$ and OBC spectrum $E_{\text{OBC}}$, all key results regarding both the local and global Lyapunov exponents are provided in the Table~\ref{table1}.

\subsection{ Self-healing phenomenon}

\begin{figure*}
\includegraphics[width=16.8cm, height=8cm]{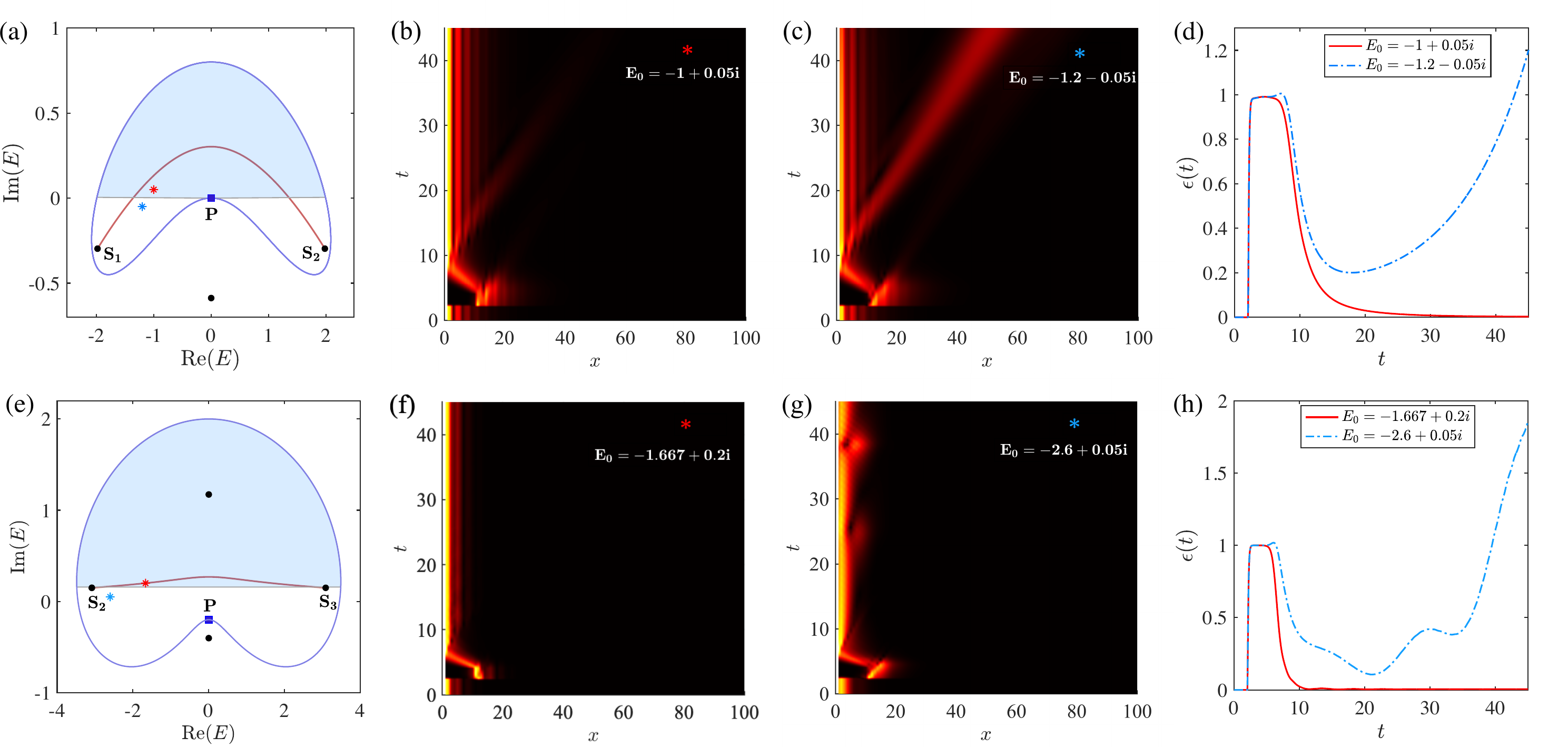}
\caption{Verification of the self-healing energy threshold predicted by Eq~\eqref{E0tot}. We consider two scenarios based on the peak velocity $v_\text{peak}$: (1) $v_\text{peak}\ne 0$ (i.e., $\text{Im}(P)>\text{Im}(S_d)$ with $S_d=S_1$ or $S_2$) [Fig.~\ref{fig4}(a)] and (2) $v_\text{peak}=0$ (i.e., $\text{Im}(P)<\text{Im}(S_d)$, with $S_d=S_2$ or $S_3$) [Fig.~\ref{fig4}(e)]. To avoid numerical errors, the spectrum is shifted along the imaginary axis by setting $\kappa=0$ in panels (a-d) and $\kappa=-0.2$ in panels (e-h). In panels (b) and (f), we show the time evolution (normalized at each $t$) of an SIBC eigenstate $\ket{\psi_0}$ whose energy $E_0$ lies within the predicted healing regime. While panels (e) and (g) show the evolution for $\ket{\psi_0}$ with $E_0$ outside that regime. Panels (d) and (h) plot $\epsilon(t)$, which quantifies how far $\ket{\psi(t)}$ deviates from the initial state $\ket{\psi_0}$. The system size is $L=600$, $\gamma=10$, and the loss range is $l=10$. A lossy potential $V=-i\gamma$ is applied in range $l$ during the time interval $[t_1=2, t_2=4)$.}
\label{fig6}
\end{figure*}

A natural application of our conclusions regarding the Lyapunov exponents is related to a phenomenon known as self-healing \cite{Longhi2022healing,Bouchal1998,McGloin2005,Broky2008,Kim2018,Antonacci2019,Thachil2024}, where classical or quantum waves can reconstruct their profiles after being disrupted or scattered by a large potential.

A recent study \cite{Longhi2022healing} revealed that non-Hermitian systems can host infinitely many skin edge modes that exhibit self-healing behavior. These modes are chosen from the eigenmodes of a non-Hermitian Hamiltonian under semi-infinite boundary conditions (SIBC), whose eigenenergies cover the interior region enclosed by the PBC (Bloch) spectrum. Moreover, Ref. \cite{Longhi2022healing} suggested that the imaginary energy part of the self-healing modes should be above the maximal imaginary part of OBC spectrum. For our model, this is equivalent to the condition $\text{Im}(E_0)>\text{max}[\text{Im}(E_\text{OBC})]=\text{Im}(O)$, where $E_0$ denotes the energy of a self-healing mode. However, upon examining our unidirectional NHSE model in Eq~\eqref{hk}, we find that this self-healing energy threshold does not always hold. 

In contrast, by utilizing our conclusions on $\lambda_\text{tot}$ derived in the preceding section, we identify the correct critical energy. A key distinction from the previous work is that our theory demonstrates that OBC eigenstates are not forbidden to be self-healing, as confirmed by the numerical results in Fig.~\ref{fig6}.

To determine whether a SIBC eigenstate is self-healing, we proceed as follows. Since it is not feasible to directly simulate a system with SIBCs, we start with a long non-Hermitian lattice under OBCs. We then prepare an initial wave packet $|\psi_0\ra$ that approximates a SIBC eigenstate: it primarily localized at the left edge and, for a short evolution time before approaching the right edge, evolves as $e^{-iHt}|\psi_0\ra=e^{-iE_0t}|\psi_0\ra$, effectively mimicking SIBCs over that short duration.
Next, we evolve the state in time and introduce a large, lossy potential $V$ during a short time interval $[t_1,t_2]$. The time-dependent Hamiltonian is:
\bea
H(t)=\left\{
\begin{aligned}
&H,  &t\in[0,t_1) \\
&H+V, &t\in[t_1,t_2]\nn\\
&H, &t\in[t_2,+\infty)
\end{aligned}.
\right.
\eea
Here, $V$ acts on the first $l$ sites near the left edge of the lattice: 
\be
V=-i\gamma\sum_{x=1}^l|x\ra\la x|.
\label{potenV}
\ee
Here, $l\ll L$ with $L$ being the system size.

To quantify how much the state $|\psi(t)\ra$ deviates from the initial state $|\psi_0\ra$, we follow the same definition in the previous paper \cite{Longhi2022healing} as:
\be
\epsilon(t)=\frac{\la\xi(t)|\xi(t)\ra}{\la\phi(t)|\phi(t)\ra},
\label{epsilon}
\ee
where $|\phi(t)\ra=e^{-iHt}|\psi_0\ra=e^{-iE_0t}|\psi_0\ra$ is the reference evolution without the perturbation $V$ and $|\xi(t)\ra=|\psi(t)\ra-|\phi(t)\ra$ represents the deviation caused by $V$.  If $\epsilon(t)$ approaches zero for $t>t_2$, it indicates that $|\psi(t)\ra$ has recovered its original profile after being disturbed, i.e. $|\psi_0\ra$ is self-healing. Conversely, if $\epsilon(t)$ grows for $t>t_2$, the initial state is not self-healing.

For our model exhibiting unidirectional NHSE, the previously proposed energy threshold for self-healing modes \cite{Longhi2022healing} was that any SIBC eigenstate with $\text{Im}(E_0)>\text{Im}(O)$ would be self-healing. However, as we will derive, the correct energy threshold is given by the Lyapunov exponent $\lambda_\text{tot}$ in short-time regime ($t<t_c$):
\be
\text{Im}(E_0)>\lambda_\text{tot}=\text{max}\left[\text{Im}(S_d),\text{Im}(P)\right].
\label{E0tot}
\ee
In general, $\text{max}\left[\text{Im}(S_d),\text{Im}(P)\right]<\text{Im}(O)$, indicating that the true healing regime is broader than previously estimated. To verify our conclusion in Eq~\eqref{E0tot}, we present two examples in Fig.~\ref{fig6}. In the first example [Fig.~\ref{fig6}(a-d), where $\lambda_\text{tot}=\text{Im}(P)>\text{Im}(S_d)$], we predict that all SIBC eigenmodes with energy $\text{Im}(E_0)>\text{Im}(P)$ (the light blue region) can self heal. For the initial state with $E_0=-1+0.05i$ (red star in the healing region), both the time evolution [Fig.~\ref{fig6}(b)] and $\epsilon(t)$ [Fig.~\ref{fig6}(d)] confirm that it is indeed self-healing. In contrast, for the state $E_0=-1.2-0.05i$ (blue star outside the healing region), $\epsilon(t)$ keeps growing over time, showing that this eigenstate cannot heal. In the second example [Figs.~\ref{fig6} (e-h)], almost the entire OBC spectrum falls within the healing region, as verified by the time-evolution of an OBC eigenstate with energy $E_0=-1.667+0.2i$ in Fig.~\ref{fig6}(f).

These results  clearly show that OBC eigenstates can also be self-healing, contradicting the earlier conclusion in Ref. \cite{Longhi2022healing}. Since OBC eigenmodes are generally easier to prepare than SIBC eigenmodes, the faithful threshold predicted by us  facilitates experimentally verifying the self-healing phenomenon on various state-of-the-art platforms.

Here, we derive the energy threshold predicted in Eq~\eqref{E0tot}. First,  
we evolve the initial state $|\psi_0\ra$, which is an eigenstate of the SIBC Hamiltonian, up to time $t_1$. Since the lattice is sufficiently long and can be approximately treated as having SIBCs, the state evolves as: $|\psi(t_1)\ra=e^{-iE_0t_1}|\psi_0\ra$. Next, we introduce a large local disruption potential $V$ acting over the range $x\in[1,l]$ during a short time interval $\Delta t=|t_2-t_1|$. The resulting state at time $t_2$ is given by:
\bea
 |\psi(t_2)\ra&=&e^{-i(H+V)\Delta t}|\psi(t_1)\ra\nn\\
 &=&\sum_{n=0}^\infty\frac{\left[-i(H+V)\Delta t\right]^n}{n!}\ket{\psi(t_1)}\nn\\
 &=&\left[e^{-iH\Delta t}+F(V,H,\Delta t)\right]\ket{\psi(t_1)}\nn\\
 &=&e^{-iE_0t_2}\ket{\psi_0}+\ket{\xi(t_2)},
\label{eq32}
\eea
where $F(V,H,\Delta t)$ contains all terms involving $V$. Terms like $VH^{n-1},V^2H^{n-2},\dots$ acting on $\ket{\psi(t_1)}$ produce a state localized near the left edge, while terms like $H^{n-1}V$ cause the disruption to spread further. However, since $V$ only acts for a short time $\Delta t$, the resulting state $\ket{\xi(t_2)}$ spreads only over a finite region. 

After time $t_2$, the Hamiltonian reverts to $H$, resulting in:
\bea
\ket{\psi(t)}&=&e^{-iH(t-t_2)}|\psi(t_2)\ra\nn\\
&=&e^{-iE_0t}|\psi_0\ra+e^{-iH(t-t_2)}|\xi(t_2)\ra\nn\\
&=&\ket{\phi(t)}+\ket{\xi(t)},
\label{phi+xi}
\eea
where $\ket{\phi(t)}$ and $\ket{\xi(t)}$ are defined as in Eq~\eqref{epsilon}. Here, $\ket{\phi(t)}$ is a stationary state under the evolution governed by the unperturbed Hamiltonian $H$, and $\ket{\xi(t)}=e^{-iH(t-t_2)}|\xi(t_2)\ra$ represents the deviation that will be analyzed. To evaluate $\ket{\xi(t)}$, we treat $|\xi(t_2)\ra$ as the initial state and evolve it under the Hamiltonian $H$. Since $|\xi(t_2)\ra$ can be interpreted as a localized wave packet near the left boundary of the system, guaranteed by the observations made in the Appendix~\ref{app:D}, our former conclusions are valid in predicting long-time behaviors of the norm of $|\xi(t_2)\ra$. According to our predictions, when $t\gg t_2$, 
\bea
\sqrt{\la \xi(t)\ket{\xi(t)}}&\sim& e^{\lambda_\text{tot}t}=e^{\text{max}\left[\text{Im}(S_d),\text{Im}(P)\right]t}.
\eea
At the same time, the norm of $\ket{\phi(t)}$ grows as $\sqrt{\la\phi(t)|\phi(t)\ra}\sim e^{\text{Im}(E_0)t}$.
Comparing these growth rates, if $\text{Im}(E_0)>\text{max}\left[\text{Im}(S_d),\text{Im}(P)\right]$, the $\ket{\phi(t)}$ term dominates in Eq~\eqref{phi+xi}, leading to a self-healing behavior. Otherwise, $\ket{\xi(t)}$ grows faster, causing $\epsilon(t)=\frac{\la\xi(t)|\xi(t)\ra}{\la\phi(t)|\phi(t)\ra}$(defined in Eq~\eqref{epsilon}) to increase with time, indicating that $\ket{\psi_{0}}$ cannot heal itself.

\subsection{The crossover timescale $t_c$}

In the above discussions, we thoroughly explored the local and global Lyapunov exponents, and identified a crossover in their behavior across two time regimes, $t\in[0, t_c]$ and $t\in[t_c,+\infty)$. Around the crossover timescale $t_c$, the particle's dynamics gradually shift from being bulk-dominated to boundary-dominated. Physically, we interpret $t_c$ as the time required for the particle or information to propagate through the lattice and crossover into a regime where boundary effects become significant.

For a complex Bloch Hamiltonian $h(k)$, its imaginary part determines the amplification or attenuation of the wavefunction amplitude, while its real part is generally considered as a generalized energy. Therefore, we define a bulk group velocity, $v_\text{bulk}(k)$, from the derivative of the real part $\text{Re}[h(k)]$ as:
\be
v_\text{bulk}(k)=\frac{d\text{Re}[h(k)]}{dk}.
\label{eqvk}
\ee
We interpret this bulk group velocity as the typical speed of information propagation, which directly determines the timescale $t_c$. Specifically, $t_c$ should correspond to the time required for the fastest propagation modes to start from $x_0=1$, reach the opposite boundary at $L$, and then return to $x_0$. Thus, a reasonable estimate for $t_c$ is:
\be
t_c=\frac{L-1}{|v_+|}+\frac{L-1}{|v_-|}= \frac{2(L-1)}{v_c},
\label{eqtc}
\ee
where $v_c=\frac{2|v_+v_-|}{|v_+|+|v_-|}$ represents the averaged velocity. Here, $v_+,v_-$ are the maximum velocities in the rightward ($v_+$) and leftward ($v_-$) directions, respectively, defined as
\bea
v_+&=&\text{max}\big[v_\text{bulk}(k)\big]=\text{max}\big[\frac{d\text{Re}(h(k))}{dk}\big]_{k\in[0,2\pi)}\nn\\
v_-&=&\text{min}\big[v_\text{bulk}(k)\big]=\text{min}\big[\frac{d\text{Re}(h(k))}{dk}\big]_{k\in[0,2\pi)}.
\label{eqv+v-}
\eea

\begin{figure}[t]
\centering
\includegraphics[width=8.4cm, height=8cm]{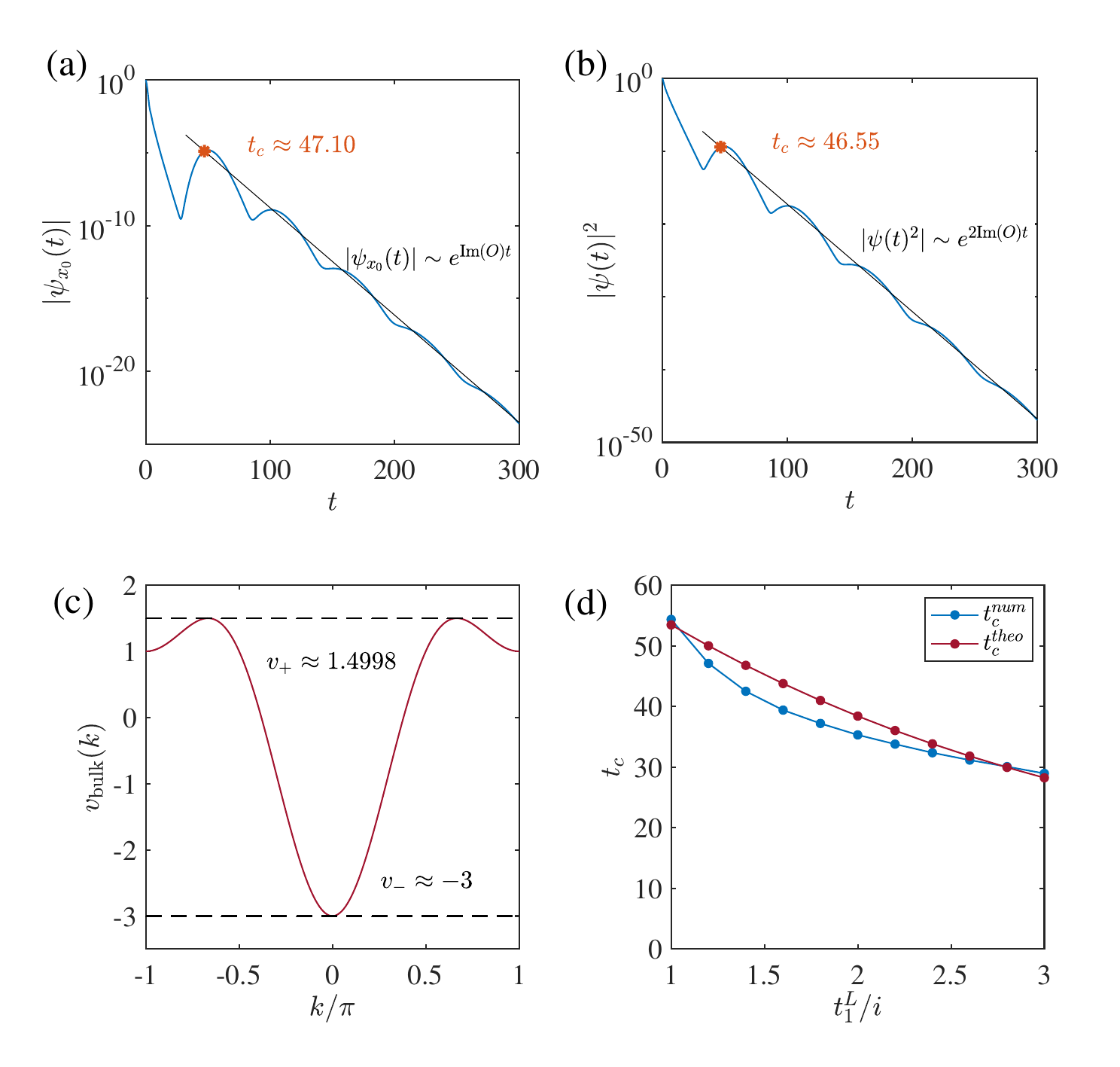}
\caption{ With parameters $t_1^L=1.2i, t_1^R=-0.8i, t_2^L=0.6i, t_2^R=0.1i$ and $\kappa=0.7$, the same as Fig.~\ref{fig2}(b) and system size $L=51$. (a)(b) The black lines show the fitting results of $|\psi_{x_0}(t)|\sim e^{\text{Im}(O)t}$ and $|\psi(t)|^2\sim e^{2\text{Im}(O)t}$ in the long time scale. (c) The maximum velocities $v_+$ and $v_-$, given by Eq~\eqref{eqv+v-}. (d) Comparison of the theoretical prediction of $t_c^\text{theo}$ by Eq~\eqref{eqtc}
and numerical results of $t_c^\text{num}$ (with $t_c$ obtained by the crossing points as shown in panels (a) and (b)), for different $t_1^L$.}
\label{fig7}
\end{figure}

As shown in Fig.~\ref{fig7}(a), we identify $t_c$ as
the first point where $|\psi_{x_0}(t)|$ begins to grow and intersects with the long-time fitting line $|\psi_{x_0}(t)|\sim
e^{\text{Im}(O)t}$. The total wavefunction norm $|\psi(t)|^2$ behaves similarly, giving $t_c =  46.55 \approx 47.10$. For the selected parameter set, we obtain the two maximum velocities as $v_+\approx 1.4998$ and $v_-\approx-3$, as shown in Fig.~\ref{fig7}(c). Using Eq~\eqref{eqtc}, these result in a predicted crossover time $t_c\approx 50.0044$. This prediction is reasonably consistent with the numerical result of $t_c \approx 47.10$. To further verify our conclusion, we vary the parameter $t_1^L$ and compare the numerically obtained $t_c^\text{num}$ with the theoretical prediction from Eq~\eqref{eqtc}, as shown in Fig.~\ref{fig7}(d). Our findings indicate that our approach can effectively predict $t_c$, despite the presence of a small discrepancy.

It should be noted that the bulk group velocity $v_+$ and $v_-$, is fundamentally different from the peak velocity $v_\text{peak}$. It is possible that $v_\text{peak}=0$ while $v_+$ and $v_-$ are non-zero. For instance, as shown in Fig.~\ref{fig4}(g), the wave peak remains pinned at the left boundary, resulting in a zero-valued $v_\text{peak}$. Nevertheless, the global Lyapunov exponent still switches from $\lambda_\text{tot}$ to $\mu_\text{tot}$ at $t_c$, which can be attributed to the non-zero bulk velocities $v_+$ and $v_-$.

\section{Discussion and conclusions}
In this work, we have derived general expressions for two kinds of Lyapunov exponents in non-Hermitian dynamics, which are summarized in Table \ref{table1}. We show that these exponents are linked to different aspects of non-Hermitian spectrums. The Lyapunov exponents before a crossover timescale can be unambiguously determined by employing the Lefschetz thimble method to find the dominant saddle point among multiple candidates, while the Lyapunov exponents after a crossover timescale are offered by the OBC spectrum. These findings reveal that boundaries play a profound role in shaping non-Hermitian real-time dynamics. From an experimental perspective, the unexpected boundary-induced dynamical phenomena found in our work are testable in various non-Hermitian platforms and can shed light on developing novel non-Hermitian techniques to manipulate wave dynamics.

\section*{Acknowledgment}

We thank Shunyu Yao for helpful discussions. This work is supported by the National Natural Science Foundation of China under Grant No. 12125405, National Key R\&D Program of China (No. 2023YFA1406702), and the Innovation Program for Quantum Science and Technology (No. 2021ZD0302502).  F.S. acknowledges supports from NSFC under Grant No.~12404189 and from the Postdoctoral Fellowship Program of CPSF under Grant No. GZB20240732.

\appendix
\renewcommand{\appendixname}{APPENDIX}
\section{\MakeUppercase{Saddle points in the complex plane}}\label{app:A}
Here, we aim to demonstrate that for a complex momentum $k=k_{r}+ik_{i}$, the extremum of $h(k)$, where $\frac{dh(k)}{dk}=0$, is always a saddle point. As a complex function, $h(k)$ satisfies Cauchy–Riemann equations: $\frac{\partial{\text{Re}h(k)}}{\partial{k_r}}=\frac{\partial{\text{Im}h(k)}}{\partial{k_i}}$ and $\frac{\partial{\text{Im}h(k)}}{\partial{k_r}}=-\frac{\partial{\text{Re}h(k)}}{\partial{k_i}}$. From these, we can derive that:
\be
\frac{\partial^2 \text{Re} h(k)}{\partial {k_r}^2}+\frac{\partial^2 \text{Re} h(k)}{\partial {k_i}^2}=0, \quad \frac{\partial^2 \text{Im} h(k)}{\partial {k_r}^2}+\frac{\partial^2 \text{Im} h(k)}{\partial {k_i}^2}=0,
\ee
 which indicates that the second derivatives in the $k_r$ and $k_i$ directions are always opposite. Therefore, at the point where $dh(k)/dk=0$, if Re$h(k)$ is a local maximum, Im$h(k)$ will be a local minimum, making this point a saddle point in the complex $k_r-k_i$ plane. 

In addition, by multiplying the two Cauchy-Riemann equations,
we obtain:
\be
\frac{\partial\text{Re}h(k)}{\partial k_r}\frac{\partial\text{Im}h(k)}{\partial k_r}+\frac{\partial \text{Re}h(k)}{\partial k_i}\frac{\partial\text{Im}h(k)}{\partial k_i}=0,
\ee
which indicates that the two gradient vectors, $\nabla\text{Re}h(k)=(\frac{\partial\text{Re}h(k)}{\partial k_r},\frac{\partial\text{Re}h(k)}{\partial k_i})$ and $\nabla\text{Im}h(k)=(\frac{\partial\text{Im}h(k)}{\partial k_r},\frac{\partial\text{Im}h(k)}{\partial k_i})$ are orthogonal to each other. Consequently, the contour lines of $\text{Re}h(k)$
and $\text{Im}h(k)$ are also orthogonal, implying that the contour lines of $\text{Re}h(k)$ (i.e., curves with constant $\text{Re}h(k)$ ) are the steepest descent/ascent lines of $\text{Im}h(k)$.

\section{\MakeUppercase{Equivalence of integrals on BZ and GBZ}}\label{app:B}

Here in this section, we aim to derive the conclusion in the Eq~\eqref{eqGBZBZ} that, integrating over the BZ is equivalent to integrating over the GBZ. As an example, we consider the case illustrated in Fig.~\ref{fig3}($b_1$-$b_4$) of the main article. For clarity, we present it again in Fig.~\ref{figGBZ} below, where we plot the GBZ (magenta line) and the BZ (dark red line) in the complex $k_r-k_i$ plane, connecting them via the paths $l_1$ and $l_2$.

\begin{figure}[H]
\centering
\includegraphics[width=8cm, height=4.8cm]{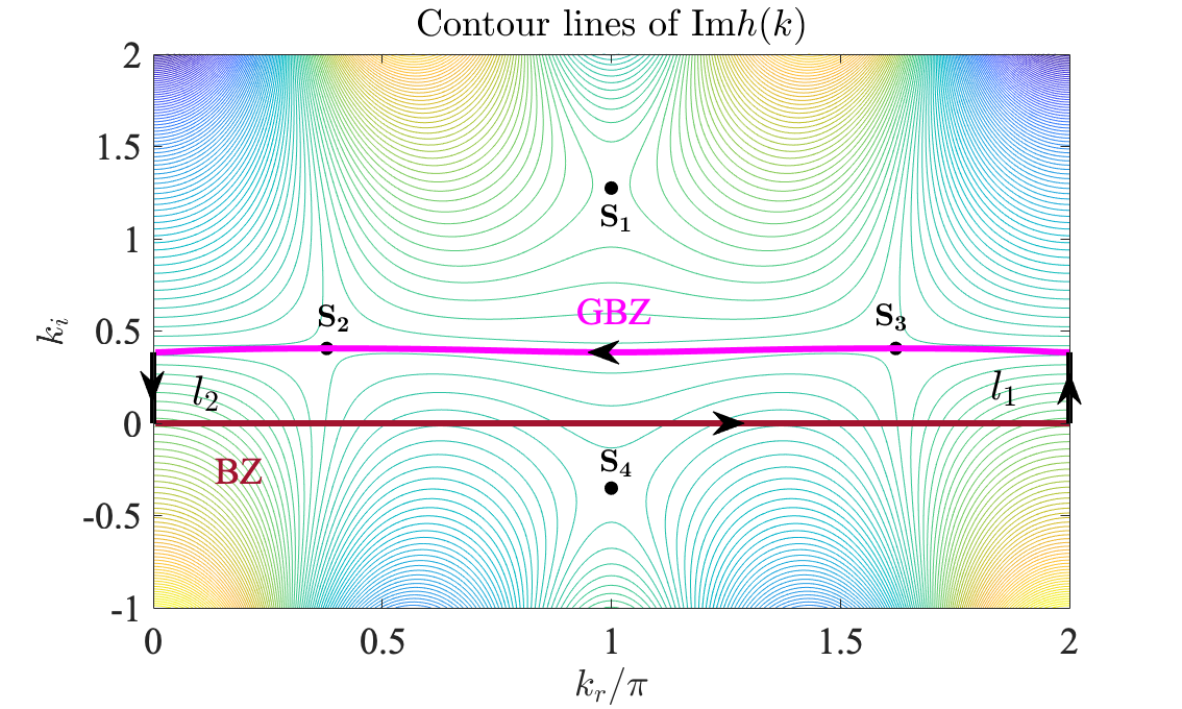}
\caption{Illustration of the closed loop $L$ formed by connecting the GBZ and BZ. With parameters $t_1^L=2.05i, t_1^R=-0.95i, t_2^L=0.85i, t_2^R=-0.15i, $ and $\kappa=0.15$, identical to those in Fig.~\ref{fig3}($e-h$).}
\label{figGBZ}
\end{figure}

By the Cauchy integral theorem, since $e^{-ih(k)t}g(k)$ is a holomorphic function within the simply connected domain enclosed by the loop $L$ (formed by $l_1\to \text{GBZ}\to l_2\to \text{BZ}$), the integral over $L$ is given by:

\be
\frac{1}{2\pi}\oint_{L}e^{-ih(k)t}g(k)dk=0
\ee

Next, it is important to note that the GBZ, like the BZ, also preserves periodicity.  Specifically, the imaginary part $k_i^\text{GBZ}$ remains unchanged if the real part $k_r^\text{GBZ}$ is shifted by $2\pi$.  Consequently, the integrals along the paths $l_1$ and $l_2$ which connect the GBZ and BZ in opposite directions (as indicated by the black arrows), satisfy:

\be
\int_{l_1}e^{-ih(k)t}g(k)dk +\int_{l_2}e^{-ih(k)t}g(k)dk=0.
\ee

By combining these results, we conclude that the summation over the GBZ ($k_r^\text{GBZ}$ ranging from $2\pi\to 0$) and BZ ($k^\text{BZ}$ ranging from $0\to 2\pi$) cancels out, leading to the equivalence between the integration over GBZ and integration over BZ. Furthermore, this equivalence extends to any curve in the $k_r-k_i$ plane that satisfies the periodicity condition $k_r \to k_r + 2\pi$. This property also justifies why the integration path can be replaced with the steepest descent curves.

\section{\MakeUppercase{Derivation of $\lambda(v_\text{peak})=\text{Im}(P)$ for non-sticky wave peaks}}\label{app:C}

Here, we explain why the maximum value $\lambda(v_\text{peak})$ in Fig.~\ref{fig5}(a) corresponds to the point $P$ on the PBC spectrum (Fig.~\ref{fig4}(a)). Specifically, we will show that $v_\text{peak}=v_P$ and hence
\bea \label{eq:vp=ImP}
\lambda(v_\text{peak})=\text{Im}(P).
\eea

To achieve this, we first show that the function $\lambda(v)=\text{Im}\left[h(k^s_d)-k^s_dv\right]$, where $\frac{dh(k)}{dk}|_{k=k^s_d}=v$, reaches its extreme value only when the dominant saddle point $k_d^s$ lives on the Brillouin zone. For brevity, we denote the dominant saddle point $k^s_d$ simply as $k^s$. Taking the derivative of $\lambda(v)$ with respect to the real number $v$ gives:
 
\bea
\frac{d\lambda(v)}{dv}&=&\text{Im}\left[\frac{dh(k^s)}{dv}-\frac{dk^s}{dv}v-k^s\right]\nn\\
&=&\text{Im}\left[\frac{dh(k^s)}{dk^s}\frac{dk^s}{dv}-\frac{dk^s}{dv}v\right]-\text{Im}(k^s)\nn\\
&=&\text{Im}\left[\left(\frac{dh(k^s)}{dk^s}-v\right)\frac{dk^s}{dv}\right]-\text{Im}(k^s)\nn\\
&=&-\text{Im}(k^s).
\eea

We have used the saddle point condition $\frac{dh(k)}{dk}|_{k=k^s_d}=v$ in the third line. Therefore, the condition of the extreme value, $\frac{d\lambda(v)}{dv}=0$, reveals that $\text{Im}(k^s)=0$, which means that the saddle point $k^s$ is purely real and lives on the Brillouin zone.

Additionally, $\frac{dh(k)}{dk}|_{k=k^s}=v$ is satisfied for a given real velocity $v$, leading to  $\frac{d\text{Im}[h(k)]}{dk}=0$ at a real momentum  $k=k^s$.  These results demonstrate that the saddle point $k^s$ for extreme $\lambda(v)$ corresponds to the extreme values of the imaginary part of the Bloch spectrum $\text{Im}[h(k)]$, like $P$ in Fig.~\ref{fig4}(a). Furthermore, the group velocities of such kind of points, $v=\frac{d\text{Re}[h(k)]}{dk}|_{k=k^s}$, provide the locations of the extreme values of the function $\lambda(v)$, and the corresponding extreme values of $\lambda(v)$ are given by $\text{Im}[h(k^s)]$. 

Generally, there are several real momentums satisfying $\frac{d\text{Im}[h(k)]}{dk}=0$. To further justify Eq.\eqref{eq:vp=ImP}, we note that only a positive velocity $v>0$ can contribute to the dynamics for the initial state localized at the left edge. Therefore, the possible extreme points of $\text{Im}[h(k)]$ should have a positive group velocity. For the model considered in the main text, the contributed point corresponds to the 
 $P$ point in Fig.~\ref{fig4}(a). We thus obtain Eq.\eqref{eq:vp=ImP} which offers the maximum value of $\lambda(v)$ in Fig.~\ref{fig5}(a).

\begin{figure*}
\includegraphics[width=16.8cm, height=8cm]{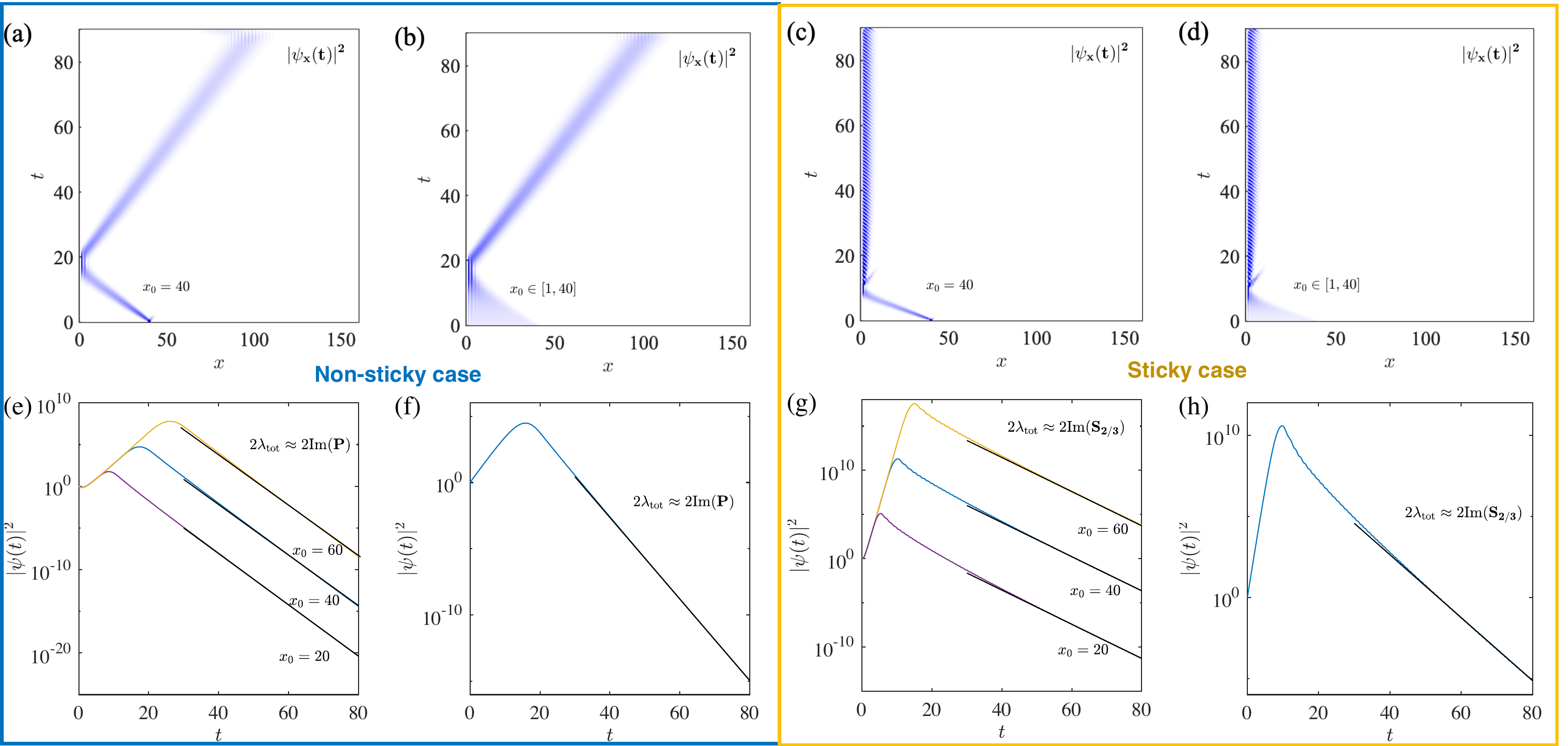}
\caption{Dynamics of the total wavefunction norm under various initial states. (a-b, e-f) Cases where the wave peak is not sticky, with parameters identical to those in Fig.~\ref{fig4}(a-d). The initial site is changed to $\ket{\psi(0)}=\ket{x_0=40}$ or $x_0=20,60$ in (a, e) and to $\ket{\psi(0)}=\frac{1}{\sqrt{l}}\sum_{i=1}^l\ket{i}$ with $l=40$ in (b, f). (c-d, g-h) Cases with a sticky wave peak, using the same parameters as in Fig.~\ref{fig4}(e-h), and considering similar initial states as in the non-sticky case. These numerical results demonstrate that even when the initial state distribution deviates from the left boundary, our conclusions regarding $\lambda_\text{tot}$ remain valid for large $t$.}
\label{figS2}
\end{figure*}

\section{\MakeUppercase{Validity for Various Initial States}}\label{app:D}
In the main article, for simplicity, we assumed that the initial state was localized at the left (skin) edge when discussing the Lyapunov exponents. Nevertheless, our conclusions for $\lambda,\mu$ and $\lambda_\text{tot},\mu_\text{tot}$ remain valid in large systems even if the initial state is not at the left edge. From Eqs~\eqref{psix0} and ~\eqref{eq18}, it is clear that the long-time exponents $\mu$ and $\mu_\text{tot}$, which are equal to the largest imaginary part of the OBC spectrum, 
are independent of $x_0$. Here, we examine the short-time behavior of the total wavefunction norm $|\psi(t)|$ for $t<t_c$ under various initial states, showing that the qualitative behavior remains consistent for sufficiently long times as well.

As shown in Fig.~\ref{figS2}, we consider two types of initial states: (1) $\ket{\psi(0)}=\ket{x_0}$, where $x_0=20, 40$ or $60$ are positions away from the left edge; (2) an evenly distributed state over the range $x\in[1,l]$, given by $\ket{\psi(0)}=\frac{1}{\sqrt{l}}\sum_{i=1}^l\ket{i}$, for both sticky and non-sticky cases. In both scenarios, as shown in Figs.~\ref{figS2}(a-d), the wave packets are first driven to the left edge by the NHSE, after which they either stick to the left edge or move to the opposite edge, depending on the parameters, consistent with our discussion in the main text. As shown in Figs.~\ref{figS2}(e-f), the total wavefunction norm initially experiences amplification when moving to the left edge, and then follows the Lyapunov exponent $\lambda_\text{tot}$, exactly as predicted in the main text. 

Therefore, for sufficiently long times $t$ in SIBC systems, our conclusions about the Lyapunov exponents remain valid for a wide range of initial states whose distribution is a finite distance from the edge at which skin modes are localized. Consequently, the self healing phenomenon also persists if the disruption $V$ is shifted further away from the left edge.

\begin{figure*}
\includegraphics[width=16.8cm, height=8cm]{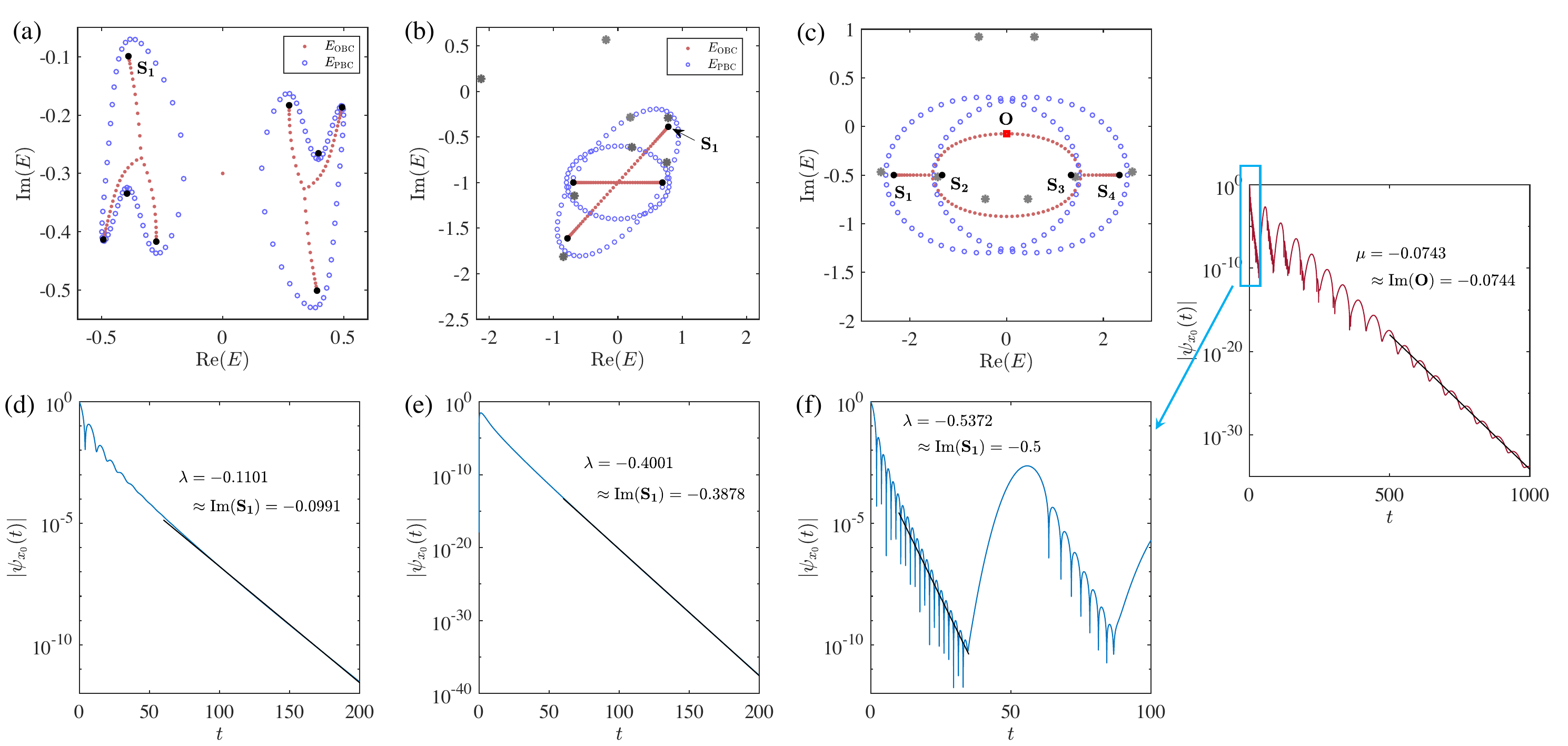}
\caption{The energy spectra, saddle points, and Lyapunov exponent $\lambda$ fittings for three two-band non-Hermitian models with system size $L=50$. (a)(d) Hamiltonian given by Eq~\eqref{2band_1} with parameters $t_1=0.1, t_2=0.4, t_3=0.1i,\gamma_1=\gamma_2=0.2$ and $\kappa=0.3$. (b)(e) Hamiltonian given by Eq~\eqref{2band_2} with parameters: $a_1=0.2, a_{-1}=0.6, b_1=0.2+0.8i, b_{-1}=0.3$, and $\kappa=2$.
The gray stars indicate the saddle points of $Q(k)=\text{det}[h(k)]$, and the black dots indicate the saddle points of $E_{1,2}(k)$.
(c)(f) Hamiltonian given by Eq~\eqref{2band_3} with parameters $t_L=1.4,t_R=0.6,V=0.5,\delta=10^{-4}$ and $\kappa=0.5$.}
\label{figS3}
\end{figure*}

\section{\MakeUppercase{Multi-band case}}\label{app"E}
Although our previous discussions have centered on a single-band model, the saddle point method naturally extends to systems with multiple bands. For an $m$-band non-Hermitian Hamiltonian $h(k)$, the Lyapunov exponents $\lambda$ and $\lambda_\text{tot}$ depend on the saddle points of all $m$ energy branches, $E_m(k)$. Below, we present several examples of different two-band scenarios to illustrate this general behavior.

First, we consider a model whose energy spectrum exhibits chiral symmetry: $E_{1,2}(k)=\pm \sqrt{\text{det}\left[h(k)\right]}=\pm \sqrt{Q(k)}$ where we define $Q(k)=\det[h(k)]$. Since $\frac{dE_{1,2}(k)}{dk}=\pm\frac{1}{2\sqrt{Q(k)}}\frac{dQ(k)}{dk}$, the saddle points of both branches $E_{1,2}(k)$ coincide with the saddle points of $Q(k)$. An example model \cite{Murakami2019} is given by:
\be
h_1(k)=\begin{pmatrix} 0 & R_+(k)\\
R_-(k) & 0
\end{pmatrix}-i\kappa\mathbb{I},
\ee
with
\be
\left\{
\begin{array}{cc}
R_+(k)=t_3e^{ik}+(t_2-\gamma_2/2)e^{-ik}+(t_1+\gamma_1/2) &  \\
R_-(k)=t_3e^{-ik}+(t_2+\gamma_2/2)e^{ik}+(t_1-\gamma_1/2)  & 
\end{array}.
\right.
\label{2band_1}
\ee
Here, we have added the constant term $-i\kappa\mathbb{I}$ to ensure that the overall system is lossy and stable. As shown in Fig.~\ref{figS3}(a), we plot the OBC and PBC energy spectra and mark the saddle points of $Q(k)$ (indicated by black dots). By applying the Lefschetz thimble method, we identify the dominant saddle point as $S_1$. According to the fittings in Fig.~\ref{figS3}(d), the Lyapunov exponent $\lambda$ agrees well with $\text{Im}(S_1)$, similar to the behavior observed in the single-band case discussed in the main text.

Then, we introduce a second model in which the saddle points of $Q(k)=\text{det}[h(k)]$ do not coincide with those of energy branches $E_m(k)$. The Hamiltonian is given by:
\be
h_2(k)=\frac{1}{2}\left[A(k)-B(k)\right]\sigma_y+\frac{1}{2}\left[A(k)+B(k)-i\kappa\right]\mathbb{I}, 
\ee
where
\be
\left\{
\begin{array}{cc}
A(k)=a_1 e^{ik}+a_{-1}e^{-ik}& \\
B(k)=b_1 e^{ik}+b_{-1}e^{-ik} & 
\end{array}.
\right.
\label{2band_2}
\ee
In this model, the two energy branches are $E_{1}(k)=A(k)-i\kappa/2, E_{2}(k)=B(k)-i\kappa/2$. In Fig.~\ref{figS3}(b), we plot the saddle points of $E_{1,2}(k)$ (black dots) and those of $Q(k)$ (light gray stars). As confirmed by the fittings in Fig.~\ref{figS3}(e), the Lyapunov exponent is determined by the dominant saddle points of $E_{1,2}(k)$, rather than by those of $Q(k)$.

Finally, we demonstrate that our conclusions apply to critical NHSE models, which exhibit size-dependent OBC spectra and GBZs. By weakly coupling two Hatano-Nelson chains with opposite skin preferences, we obtain the Hamiltonian\cite{cNHSE2020}:
\be
h_3(k)=\begin{pmatrix} t_Le^{ik}+t_Re^{-ik}+V & \delta\\
\delta & t_Re^{ik}+t_Le^{-ik}-V
\end{pmatrix}
\label{2band_3}
\ee
where $\delta\ll 1$. As shwon in Fig.~\ref{figS3}(c), the saddle points of $E_{1,2}(k)$ (black dots) do not coincide with those of $Q(k)$ (gray stars). Moreover, similar the the cases presented in the main article, the imaginary parts $\text{Im}(S_{1\sim 4})$ differ from $\text{Im}(O)$, leading to a crossover in the Lyapunov exponent $\lambda$, as verified in Fig.~\ref{figS3}(f) and its inset.

\section{\MakeUppercase{Construction of of the SIBC eigenstates}}\label{app:F}
In the preceding discussions of the self-healing phenomenon (Section III. B), the healing modes $\ket{\psi_0}$ correspond to the eigenmodes of the non-Hermitian Hamiltonian under semi-infinite boundary conditions (SIBC), with energies $E_0$ that satisfy Eq~\eqref{E0tot}. As shown in Fig.~\ref{fig6}, we selected several initial states with different $E_0$ values and verified whether they recover their original profiles after being scattered by a large potential.  However, constructing such an SIBC eigenstate in a finite-size OBC system is not straightforward. Below, we outline how to construct these SIBC eigenstates numerically.

First, consider the Hamiltonian in Eq~\eqref{hk} under SIBC, meaning the system has a single left boundary but extends infinitely to the right. A general eigenstate $\ket{\psi_0}$ can then be written as a linear combination:
\be 
\psi_0(x)=c_1\beta_1^x+c_2\beta_2^x+c_3\beta_3^x+c_4\beta_4^x,
\ee
where $|\beta_1|\le|\beta_2|\cdots\le|\beta_4|$ are the roots of 
\be
h(\beta)=t_1^L\beta+t_1^R\beta^{-1}+t_2^L\beta^2+t_2^R\beta^{-2}=E_0,\text{ with }\beta=e^{ik}.
\ee
As shown in Fig.~\ref{fig6}(a) and \ref{fig6}(e), the SIBC eigenenergy $E_0$ lies within the region enclosed by the PBC spectrum, giving the winding number
\be
W_{\text{BZ},E_0}=\frac{1}{2\pi}\oint_\text{BZ}\frac{d}{d\beta}\arg{[h(\beta)-E_0]}d\beta .
\ee
According to the Cauchy's argument principle, we have $W_{\text{BZ},E_0}=Z-P$, where $Z$ and $P$ are the numbers of zeros (i.e., $\beta_1\sim \beta_4$) and poles of $h(\beta)-E_0$ inside the BZ ($|\beta|=1$), respectively. As shown in Fig. \ref{fig6}, we find $W_{\text{BZ},E_0}=1$ in our case. Since $h(\beta)$ is a Laurent polynomial from $\beta^{-2}$, there is a second-order pole at $\beta=0$, i.e., $P=2$. This leads to $Z=P+1=3$. Hence, three roots ($\beta_{1,2,3}$) lie inside the BZ, while the remaining root must satisfy $|\beta_4|>1$.
As a result, the corresponding coefficient $c_4$ must be set to zero to prevent $\psi_0(x)$ from diverging as $x\rightarrow \infty$. We therefore have three unknown coefficients, $c_1,c_2$ and $c_3$.

Next, we apply two boundary conditions at the left edge:
\bea
\begin{cases}
-i\kappa\psi_{x_0}(1)+t_1^L\psi_{x_0}(2)+t_2^L\psi_{x_0}(3)=E_0\psi_{x_0}(1)\nn\\
t_1^R\psi_{x_0}(1)-i\kappa\psi_{x_0}(2)+t_1^L\psi_{x_0}(3)+t_2^L\psi_{x_0}(4)=E_0\psi_{x_0}(2),
\end{cases}
\eea
and then impose the normalization condition
$\sum_{x=1}^{\infty}|\psi_{x_0}(x)|^2\rightarrow1$.
Solving these three constraints yields the coefficients $c_1,c_2$ and $c_3$. In this way, we obtain a properly normalized SIBC eigenstate $\ket{\psi_{x_0}}$ with target energy $E_0$. For additional approaches to constructing SIBC eigenstates on a finite lattice, see Ref.\cite{Longhi2021Selective} and the supplemental material of Ref.\cite{Longhi2022healing}.

\section{\MakeUppercase{Additional note}}
1. The saddle point method fails when the hoppings are entirely unidirectional (e.g. $h(k)=e^{ik}$). In this case, the GBZ is no longer a closed loop but instead collapses to 0 or extends to infinity, and the OBC spectrum reduces to a single point. Consequently, there are no distinct eigenvalues to interfere and produce a Lyapunov exponent determined by the saddle point.

2. We should emphasize that if there exists a saddle point $S_\sigma$ whose imaginary part $\text{Im}(S_\sigma)$ exceeds the largest imaginary part of the OBC eigenvalues $\text{Im}(O)$, then this saddle point cannot be the dominant saddle point.
In other words, its contribution coefficient must vanishi, $n_\sigma=0$. To see this, we consider $|\psi_{x_0}(t)|$: 
\bea
\left|\psi_{x_0}(t)\right|&=&\left|\la x_0|e^{-iHt}|x_0\ra\right|=\sum_n\left|e^{-iE_nt}\la x_0|\psi_n^R\ra\la\psi_n^L|x_0\ra\right|\nn\\
&\le& e^{\text{Im}(O)t}\sum_n\left|\la x_0|\psi_n^R\ra\la\psi_n^L|x_0\ra\right|, 
\eea
and $|\psi(t)|^2$:
\bea
\sum_x|\psi(x,t)|^2&=&\sum_{x,n}e^{2\text{Im}(E_n)t}|\la x|\psi_{n}^R\ra\la\psi_{n}^L|x_0\ra|^2,\nn\\
&\le&e^{2\text{Im}(O)t}|\sum_{x,n}\la x|\psi_{n}^R\ra\la\psi_{n}^L|x_0\ra|^2.
\eea
From these inequalities, we conclude that the Lyapunov exponents $\lambda,\mu$ and $\lambda_\text{tot},\mu_\text{tot}$ are bounded above by $\text{Im}(O)$. Hence, it is not possible for a saddle point $S_\sigma$ with $\text{Im}(S_\sigma)>\text{Im}(O)$ to contribute.

3. Moreover, even if a saddle point's imaginary part is less than $\text{Im}(O)$, it does not necessarily contribute to the Lyapunov exponent. To confirm whether such a saddle point actually contributes, we must apply the Lefschetz method and determine its coefficient $n_\sigma$.

\bibliography{edgedynamic} 

\end{document}